\documentclass[aps,prl,preprintnumbers,twocolumn,superscriptaddress,nofootinbib]{revtex4-1}

\usepackage{amssymb}
\usepackage{graphicx}
\usepackage{amsmath}
\usepackage{extpfeil}
\usepackage{subfigure}
\usepackage{multirow}
\usepackage{setspace}
\usepackage{verbatim}
\usepackage{slashed}
\usepackage{float}
\usepackage{color}
\usepackage{ulem}
\usepackage[utf8]{inputenc}
\usepackage[table,xcdraw]{xcolor}
\usepackage{makecell}
\usepackage[colorlinks,citecolor=blue]{hyperref}

\usepackage[T1]{fontenc} 
\usepackage[utf8]{inputenc} 
\usepackage{times}

\begin{document}
	
	\title{Resonant Forbidden CP Asymmetry from  Soft Leptons}

	\author{Shinya Kanemura}
	\email{kanemu@het.phys.sci.osaka-u.ac.jp}
	\affiliation{Department of Physics, The  University of Osaka, Toyonaka, Osaka 560-0043, Japan}

	\author{Shao-Ping Li}
	\email{lisp@het.phys.sci.osaka-u.ac.jp}
	\affiliation{Department of Physics, The  University of Osaka, Toyonaka, Osaka 560-0043, Japan}
	
	\preprint{OU-HET-1238}

	\begin{abstract}
		To explain the baryon asymmetry in the early universe via leptogenesis, quantum corrections to new  particles are commonly invoked to generate  the necessary CP asymmetry.
		We  demonstrate, however, that a large CP asymmetry    can already arise from  Standard Model leptons. The mechanism relies  on    resummation of soft leptons   at finite temperatures. The CP asymmetry, which is kinematically  forbidden in vacuum,  can be resonantly enhanced from thermally resummed leptons   by  seven orders of magnitude. Contrary to the resonance from exotic particles, 
		we show that  the    resonant  enhancement from soft leptons is protected by    controlled    widths   under    finite-temperature perturbation theory.    We quantify such CP asymmetries in  leptogenesis with secluded flavor effects and comment on  the significance  and application. The mechanism exploits the maximal role of leptons themselves, featuring low-scale leptogenesis, minimal  model buildings and  dark matter cogenesis. 
	\end{abstract}

	\maketitle
	
	\textit{\textbf{Introduction.}} The baryon  asymmetry in the universe (BAU) remains a fundamental  question in the Standard Model (SM) of particle physics and cosmology~\cite{Planck:2018vyg}. 
	Baryogenesis via leptogenesis is a simple mechanism to create a baryon asymmetry in  the early universe~\cite{Fukugita:1986hr,Luty:1992un,Covi:1996wh,Pilaftsis:1997jf}. A common ingredient of     CP asymmetries  necessary for the BAU problem is the  onshell cuts  arising from    quantum loop diagrams of mixed particles.
	This underlies current leptogenesis, where  onshell cuts are  present in  loop diagrams of   flavored  neutrinos~\cite{Giudice:2003jh,Buchmuller:2004nz,Davidson:2008bu,Fong:2012buy} or  mixed scalars~\cite{Covi:1996fm,Ma:1998dx,Dick:1999je,Murayama:2002je}.

	While the SM leptons are the basic container for lepton-baryon processing via  sphaleron at high temperatures~\cite{Kuzmin:1985mm},  the requisite amount of  CP asymmetries  for sphaleron processing insofar  is not induced by  lepton mixing itself but is instead generated  by   new  particle mixing.  
	Given that  new physics (NP) interactions  in  leptogenesis   unavoidably induce certain mixing for   lepton flavors,  onshell cuts from lepton loop diagrams  may also generate CP asymmetries.   In vacuum,  however,    there is no kinetic cut  making   particles onshell from  lepton loop diagrams, since   the loop particles  are massive while the  SM lepton doublets are massless. 
	
	Nevertheless, particles in a thermal plasma are of  statistical nature. Loop particles being onshell can  correspond not only to emission processes but also to absorption from the plasma, with the   probability proportional to distribution functions~\cite{Weldon:1983jn}.  Such  plasma-induced effects have been considered  in loop diagrams of   neutrinos and scalars by 
	using the Schwinger–Keldysh  Closed-Time-Path (CTP) formalism~\cite{Chou:1984es,Calzetta:1986cq,Berges:2004yj}, such as   unflavored leptogenesis~\cite{Garny:2009qn,Beneke:2010wd,Garbrecht:2010sz,Garny:2010nj}, flavored leptogenesis~\cite{Beneke:2010dz,Drewes:2012ma}, resonant leptogenesis~\cite{Garbrecht:2011aw,Garny:2011hg,BhupalDev:2014oar} and scalar decay~\cite{Frossard:2012pc,Garbrecht:2012qv,Hambye:2016sby}. Noticeably,  some of these plasma-induced CP asymmetries  can work as the only CP-violating source at finite temperatures,  which is otherwise forbidden in vacuum.
	
	While plasma-induced CP asymmetries may already   arise from   leptons at finite temperatures, 
	there is   a dearth of investigations into how such CP asymmetries arise from    leptons, and particularly to what extent  the CP asymmetry  can be generated.   One   difficulty   comes from applying the   Kadanoff-Baym (KB)  mixing equations~\cite{Prokopec:2003pj,Prokopec:2004ic} to SM lepton flavors.   It is known that the unmixed KB equation  is much more difficult to manage than the conventional Boltzmann equation used in most leptogenesis. The intrinsic complexity of lepton KB mixing equations further makes the generation and evolution of CP asymmetries  more challenging to compute. Such mixed  KB   equations   were   considered  in Ref.~\cite{Garbrecht:2012pq}, where    traditional flavored leptogenesis~\cite{Beneke:2010dz} can induce a net lepton asymmetry with a noticeable  resonant enhancement. Nevertheless, the lepton-asymmetry generation   occurs far above the electroweak scale, and the  technical treatment of mixed  KB   equations  
	 makes the  resonant enhancement  difficult  to trace.
	
	In this Letter,  we   introduce the concept of  forbidden CP asymmetries from soft-lepton mixing  and  provide a simple approach to  calculate the forbidden CP asymmetries.  This approach circumvents   the lepton KB mixing equations developed in Refs.~\cite{Beneke:2010dz,Garbrecht:2012pq}  and  focuses on  the accompanying asymmetries in the NP sector. While being  indirect,  the approach offers  a simple picture of   how the CP asymmetry is generated after resumming  leptons at finite temperatures and  how resonant enhancement   appears. It also allows   easy implementation and generalization in leptogenesis.  
	The resonant forbidden CP asymmetries  can either work  as a new mechanism for leptogenesis where  no vacuum cuts exist, or contribute as an irreducible source  to   leptogenesis where  cuts  already exist  in vacuum.
	
\begin{figure}
		\centering
		\includegraphics[width=0.485\textwidth]{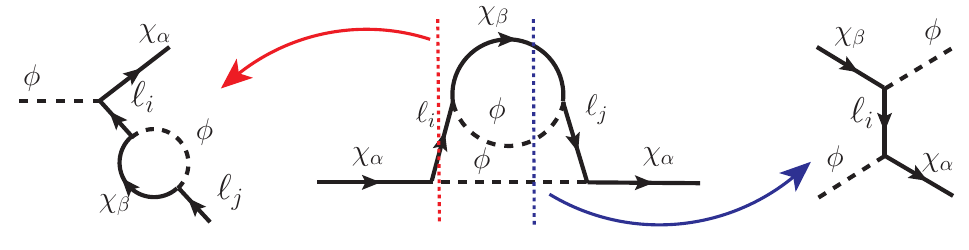} 
		\caption{\label{fig:lep-cut_Smatrix} Left: an  $S$-matrix glimpse of the resonant forbidden CP asymmetry from the  cuts  of the lepton self-energy diagram.  Middle: a two-loop diagram for CP violation in the CTP formalism, where the red cutting line has a \textit{qualitative} correspondence to the left diagram. Right: a $t$-channel scattering process mediated by massless leptons and  induced by the blue   cutting line.
		}
	\end{figure}

	\textit{\textbf{The $\boldsymbol S$-matrix picture.}}  Let us first present a simple picture of how  CP  asymmetries arise from thermally corrected leptons, and get a  glimpse of    the  resonantly enhanced forbidden CP asymmetry. In the $S$-matrix formalism,  an onshell cut from SM lepton self-energy   can be described by the left diagram in  Fig.~\ref{fig:lep-cut_Smatrix}. Consider a prototype  of  Yukawa interactions in leptogenesis:  $y'_{i\alpha}\bar \ell_i \tilde{\phi}P_R\chi_{j}+\rm h.c.$, with    $\ell$ the  SM leptons and  $\tilde \phi= i \sigma_2 \phi^*$ ($P_R\chi_j= (1+\gamma_5) \chi_j/2$)   a  gauge $SU(2)_L$   scalar  doublet (flavored fermion singlets).  
	For massive $\phi,\chi$ and massless $\ell$, 
	no kinetic phase (onshell $\phi$ and $\chi_\beta$) can arise  from the loop diagram in vacuum.  At finite temperatures,  however, 
	 particles $\phi, \chi_\beta$ in the lepton loop diagram have certain probability to be onshell,   which depends  on the statistic distribution functions. With the standard calculation of the $S$-matrix amplitude between   $\phi \to \chi_\alpha+\ell_j$ and its CP-conjugated process,  a $\chi_\alpha$ asymmetry  is generated at the order shown  on the left   of  Fig.~\ref{fig:lep-cut_Smatrix}. 
	
	When computing the   decay and   $t$-scattering channels   mediated by the massless propagator $\ell_i$,  we will encounter  infrared (IR) divergence in the limit $p_\ell \to 0$.  Such IR divergence arises from  massless leptons at high temperatures, which usually indicates the breakdown of perturbation expansion. One known technique to cure this IR divergence is the Hard-Thermal-Loop resummation~\cite{Braaten:1989mz,Carrington:1997sq,Bellac2000},   where     propagation  with soft momentum  $p\sim g T$ or $p_0^2-|\vec p|^2\sim g^2 T^2$ should be resummed over  one-particle-irreducible (1PI) self-energy diagrams.  Here $g$ denotes the weak coupling in the theory and $T$ the plasma temperature.  
Treating the resummation of leptons by inserting the  thermal masses with  the   modified dispersion relation $p^2\sim \tilde{m}^2$ into the propagator $\ell_i$ and the external state $\ell_j$, we will obtain an asymmetry of the  $\chi_\alpha$  number density scaling as
	\begin{align}\label{eq:Y-asymmetry}
		n_{\chi_\alpha}-n_{\bar \chi_{\alpha}}\propto \mathcal{O}(y^{\prime 4})\times \frac{m_\phi^2}{\tilde{m}_j^2-\tilde{m}_i^2}\,.
	\end{align} 
	Such scaling   is  generically expected from  self-energy cuts in conventional leptogenesis~\cite{Giudice:2003jh,Buchmuller:2004nz,Davidson:2008bu,Fong:2012buy}. Here the asymmetry is generated at   the quartic order of Yukawa couplings,  the   mass $m_\phi$ is introduced by dimensional arguments,  and~\cite{Weldon:1982bn} 
	\begin{align}\label{eq:m_L}
		\tilde{m}_{i}^2=\left(\frac{3}{32}g_2^2+\frac{1}{32}g_1^2+\frac{1}{16}y_{i}^2\right)T^2\,,
	\end{align}
	denotes the squared thermal mass of lepton flavor $i$ with gauge couplings $g_2, g_1$ and the lepton Yukawa coupling $y_{i}$. The difference   in the denominator of Eq.~\eqref{eq:Y-asymmetry} cancels the dependence on gauge couplings, giving rise to  a maximal enhancement  $16/(y_{\mu}^2-y_e^2)\gtrsim 4.4\times 10^7$ at $T\lesssim  m_\phi$.
	The enhancement mimics the mass degeneracy between two neutrinos or scalars in resonant leptogenesis~\cite{Pilaftsis:2003gt,Pilaftsis:2005rv}. 
	
	The  evolution of the CP asymmetry generated from the left diagram of Fig.~\ref{fig:lep-cut_Smatrix} may be quantified by   Boltzmann equations with  $S$-matrix amplitudes.  However,  
	since  the CP asymmetry is a thermal quantum-statistical effect,  using  the the semi-classical Boltzmann equations can lead to inconsistent conclusions. This can be inferred from the   two-loop diagram in the middle of Fig.~\ref{fig:lep-cut_Smatrix}, where    decay on the left  of  Fig.~\ref{fig:lep-cut_Smatrix} originates from the red cutting  line and   $t$-channel scattering on the right   is induced by the blue cutting line. The scattering channel inherited from the middle diagram   contains both   onshell and offshell lepton propagation. The CP asymmetry from the onshell mode   is essential to guarantee   CPT invariance and unitarity~\cite{Giudice:2003jh,Frossard:2012pc}. This cancellation avoids the  double-counting issue which,  in contrast,  occurs  in   Boltzmann equations if one only considers   decay and inverse decay~\cite{Kolb:1979qa}. 
	
	For practical calculations, the optical theorem  may be formulated in  finite-temperature field  theory so  that  one can insert proper  Green's functions in the left diagram to estimate the forbidden CP asymmetry~\cite{Kobes:1990ua,Garny:2010nj,Frossard:2012pc,Hambye:2016sby,Li:2020ner,Li:2021tlv}. Nevertheless,  the optical theorem in finite-temperature field  theory~\cite{Kobes:1985kc,Kobes:1986za} does not always provide  simpler ways to calculate the CP asymmetry, and these calculations could still   miss  the contribution from   onshell scattering that has a canceling  effect on the decay-induced CP asymmetry.  An important feature in this respect is that  the out-of-equilibrium condition provided  by the decay product $\chi_\alpha$ is not sufficient   to generate  a nonzero forbidden CP asymmetry, as will be shown below by
	using  the  two-loop diagram in the CTP formalism.
	
\begin{figure}[t]
		\centering
		\includegraphics[scale=0.52]{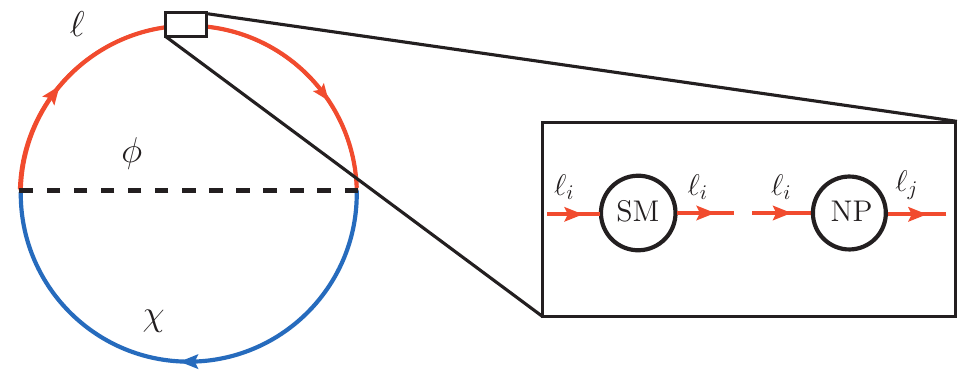} 
		\caption{\label{fig:2PI} The 2PI effective action $\Gamma_2$ at two-loop order in the CTP formalism. The propagators in the closed loop are in their \textit{full} forms. Particularly, the resummed lepton propagator includes   1PI diagrams from the SM flavor-conserving  gauge and Yukawa interactions, and the NP flavor-violating Yukawa interactions. 
			}
	\end{figure}

	\textit{\textbf{CP asymmetry in the CTP formalism.}} The resummation of  soft leptons discussed above is  motivated not only by IR divergence, but also by the built-in procedure  in the CTP formalism.  The collision rates in the KB equation  come from the functional derivative of the two-particle-irreducible (2PI) effective action~\cite{Calzetta:1986cq,Berges:2004yj}, which is  formed by closed loops with full   propagators. As shown in Fig.~\ref{fig:2PI},   the resummation of leptons includes  the 1PI diagrams from flavor-conserving  SM  and flavor-changing NP contributions.
	
For simplicity, we consider two $\chi$ flavors. In the basis where the lepton Yukawa and $\chi$-mass matrices are diagonal, the $y'$-Yukawa interactions   induce lepton flavor mixing.  
	To  quantify the impacts of  the forbidden CP asymmetry in leptogenesis,  we consider an example where the total lepton number $L_{\rm SM}+L_{\chi}$ is conserved or approximately  conserved during the main epoch of leptogenesis. 
	In this example,  leptogenesis is triggered via the  $B+L_{\rm SM}$ violating but $B-L_{\rm SM}$ conserving sphaleron process, once  an asymmetry is generated and  secluded in the $\chi$ sector  satisfying   $B-L_{\rm SM}\approx L_\chi$.  
	This realization   mimics  the idea of \textit{hiding lepton (or baryon) numbers in the dark sector}, where dark or secluded flavor effects can retain a net  asymmetry  from  lepton- (baryon-) number conserving processes~\cite{Akhmedov:1998qx,Dick:1999je,Davoudiasl:2010am,Cheung:2011if,Elor:2018twp}.
	Consider  the  situation where  $\chi_1$ is out of equilibrium while $\chi_2$ keeps in thermal equilibrium with the SM plasma.   The equilibrium condition of $\chi_2$ indicates that no CP asymmetry in $\chi_2$ will be generated, and 
	the final baryon asymmetry is determined by the unmixed KB equation of $\chi_1$ through the relation~\cite{Harvey:1990qw}
	\begin{align}\label{eq:YB}
		Y_{B}\equiv \frac{n_{B}- n_{\bar B}}{s}\approx 0.35 Y_{\chi_1}\,,
	\end{align}	 
	where $Y_{\chi_1}\equiv (n_{\chi_1}- n_{\bar \chi_1})/s_{}$  with $s\approx 46.83 T^3$   the SM entropy density, and $Y_B\approx 8.75\times 10^{-11}$ is the observed baryon asymmetry in the universe~\cite{Planck:2018vyg}.
	
	In a spatially homogeneous  background, the KB equation of  $\chi_1$ can be formally written as~\cite{SM}
	\begin{align}\label{eq:KB}
		\frac{d (sY_{\chi_1})}{d t}=\mathcal{W}+\mathcal{S}_{\rm CP}\,,
	\end{align}
	where  $\mathcal{W}$ denotes the washout rate and $\mathcal{S}_{\rm CP}$ the CP-violating source
	\begin{align}\label{eq:S_CP}
		\mathcal{S}_{\rm CP}=\frac{1}{2}\int \frac{d^4 p}{(2\pi)^4} \text{Tr}\left[i\slashed\Sigma_{\chi_1}^> i\slashed S_{\chi_1}^<-i\slashed\Sigma_{\chi_1}^<i \slashed S_{\chi_1}^>\right],
	\end{align}
with   $i \slashed S_{\chi_1}^{<,>}, i\slashed\Sigma_{\chi_1}^{>,<}$  the  $\chi_1$ Wightman functions and self-energy amplitudes, respectively, and $\rm Tr$ the   Dirac trace.  In full thermal equilibrium, the Kubo–Martin–Schwinger relations hold: $\slashed\Sigma_{\chi_1}^>(p)=-e^{p_0/T}\slashed\Sigma_{\chi_1}^< (p), \slashed S_{\chi_1}^>(p)=-e^{p_0/T}\slashed S^<_{\chi_1}(p)$, such that the source term   vanishes and no $\chi_1$ asymmetry will be  generated. 
	The leading contribution to the washout rate $\mathcal{W}$ is determined by   one-loop self-energy diagrams of $\chi_1$ without resummation, while the forbidden  CP asymmetry of  $\chi_1$ comes from the middle diagram of Fig.~\ref{fig:lep-cut_Smatrix}.  The calculation of $\mathcal{S}_{\rm CP}$ constitutes one of the main difficulties in the KB equation, but we find that it can be greatly simplified after exploiting   the integral symmetries and   reasonable approximations. The final result succinctly reads~\cite{SM}
	\begin{align}\label{eq:nonthermal-chi1-rate_1}
		\mathcal{S}_{\rm CP}&=\frac{\text{Im}(y^{\prime 4}) m_\phi^4}{256\pi^4(\tilde{m}_j^2-\tilde{m}_i^2)}\mathcal{I}\,,
	\end{align}
	where  $\text{Im}(y^{\prime 4})\equiv \text{Im}[y'_{i1}y'^{*}_{j1}y'^{*}_{i2}y'_{j2}]$   with $i\neq j$   summed over three lepton flavors.   

	The function $\mathcal{I}$ represents the statistic nature of the forbidden CP asymmetry:
	\begin{align}\label{eq:I}
		\mathcal{I}&= \int_0^\infty  \frac{dp_\ell}{p_\ell} 
		\int^\infty_{p_\ell+m_\phi^2/(4p_\ell)} dE_\phi \int^\infty_{p_\ell+m_\phi^2/(4p_\ell)}  dE'_\phi 
		\nonumber\\[0.2cm]
		&\times f_\phi^{\rm eq}(E_\phi)\delta f_\phi(E'_\phi)\left[f^{\rm eq}_{\chi_2}(E_\phi'-p_\ell)+f^{\rm eq}_\ell(p_{\ell})-1\right],
	\end{align}
	where  $\delta f_\phi \equiv f_\phi-f_\phi^{\rm eq}$, $E_\phi^{(\prime)}$ is   the scalar energy in the outer (inner) loop,  and   the thermal distribution functions for  $\chi_2,\ell$ are used. 
We can   infer from  Eq.~\eqref{eq:I} the underlying nature of  the forbidden CP asymmetry:
  the generation rate of forbidden CP asymmetries must be scaled  with    the distribution functions of  the inner loop particles.  This is understood if we  omit   the  statistics of  $\phi$ in the inner loop: $f_\phi^{(\rm eq)}(E'_\phi)=0$, which gives rise  to $\mathcal{S}_{\rm CP}=0$. Thus, we quantitatively confirm  the expectation outlined at the beginning:     the probability of  the forbidden CP asymmetry is   proportional to distribution functions of the inner onshell particles.

	Using Eq.~\eqref{eq:m_L}, we can estimate the CP-violating source at $T\simeq m_\phi$ as
	$\mathcal{S}_{\rm CP}\propto \text{Im}(y^{\prime 4})/(y_j^2-y_i^2)$,
	where  the  flavor-universal gauge couplings are canceled and the  survived lepton Yukawa couplings   cause an enhancement.  Note that  thermal widths of leptons are not included in Eq.~\eqref{eq:nonthermal-chi1-rate_1}. Actually,  the quasiparticle nature of a thermal lepton flavor $i$ not only has a flavor-dependent pole: $p_{0}^2-|\vec p|^2\approx 2\tilde{m}_i^2$~\cite{Weldon:1982bn, Kiessig:2010pr,Drewes:2013iaa,Li:2023ewv}, but  also predicts a thermal width at  $\mathcal{O}(g^4)$~\cite{SM}.  Since $g^4\gg y_i^2$ for all lepton flavors, inserting the width into the denominator of Eq.~\eqref{eq:nonthermal-chi1-rate_1} seems to  reduce the strong resonance.   
	However,  the insertion  is not   a consistent treatment under finite-temperature perturbation theory.  The underlying reason is that the leading thermal widths are at the same order in gauge couplings for all lepton flavors, which dictates that the thermal widths at a given order should be treated universally for all leptons. In the following, we present a simple way to see this, and give a demonstration with the Cauchy's residue theorem   in~\cite{SM}.

	Tracing  the  scaling  $1/(y_j^2-y_i^2)$ can be simplified  by decomposing  the two-loop amplitudes $i\slashed\Sigma_{\chi_1}^{>,<}$ into    products of   resummed lepton retarded propagator $\slashed{S}_{\ell_i}^R$, Wightman function $\slashed{S}_{\ell_j}^{<}$ and  the inner-loop amplitudes $\slashed{\Sigma}_{\ell_{ij}}^{ab}$ with  $a, b$ the CTP indices. Factoring out the spin structures,   we can write the unslashed propagators as~\cite{Braaten:1990wp}
	\begin{align}
		{S}^R_{\ell_i}\propto  \frac{1}{\text{Re}\Delta_i+i \text{Im}\Delta_i},~
		{S}^{<}_{\ell_j} \propto  \frac{\text{Im}\Delta_j}{(\text{Re}\Delta_j)^2+(\text{Im}\Delta_j)^2}\,.\label{eq:SR<>}
	\end{align}
	The real part $\text{Re}\Delta$ determines the modified dispersion relation while the imaginary part $\text{Im}\Delta$ represents a finite width. At the  pole $p^2\approx 2\tilde{m}_j^2$,  the leading-order dependence of both $\text{Im}\Delta_i$ and $\text{Im}\Delta_j$ on the flavor-universal gauge couplings is at $\mathcal{O}(g^4)$, such that $S^R_{\ell_i}(p)  S^{<}_{\ell_j}(p)\propto   \delta(p^2-2\tilde{m}_j^2)/(\tilde{m}_j^2-\tilde{m}_i^2)$ holds up to  $\mathcal{O}(g^2)$.  This is different from vacuum mass degeneracy   invoked  in   resonant leptogenesis where two neutrino/scalar flavors having vacuum mass degeneracy do not have guaranteed decay widths at the same   order of couplings.  
	Going to $\mathcal{O}(g^4)$, we may write $S^{<}_{\ell_j}(p) \propto \delta[p^2-2\tilde{m}_j^2- g^4 q_1 -i g^4 q_2]$, with $q_{1,2}$ the functions of momentum and temperature~\cite{SM}. It is straightforward to  check $S^R_{\ell_i}(p)  S^{<}_{\ell_j}(p)\propto   \delta(p^2-2\tilde{m}_j^2- g^4 q_1 -i g^4 q_2)/(\tilde{m}_j^2-\tilde{m}_i^2)$, where the $\mathcal{O}(g^4)$ terms from  $S^R_{\ell_i}$ and  $S^{<}_{\ell_j}$ cancel in  the denominator.   We can further go to higher orders of gauge couplings   and repeat the above perturbation analysis.  Generically, all the flavor-universal terms will be canceled in the product  $S^R_{\ell_i}  S^{<}_{\ell_j}$, leaving  $g^2 y_i^2$ and $g^2 y_j^2$ as the next-to-leading-order terms in the denominator of Eq.~\eqref{eq:nonthermal-chi1-rate_1}. 	
	
\begin{figure}[t]
		\centering
		\includegraphics[scale=0.3]{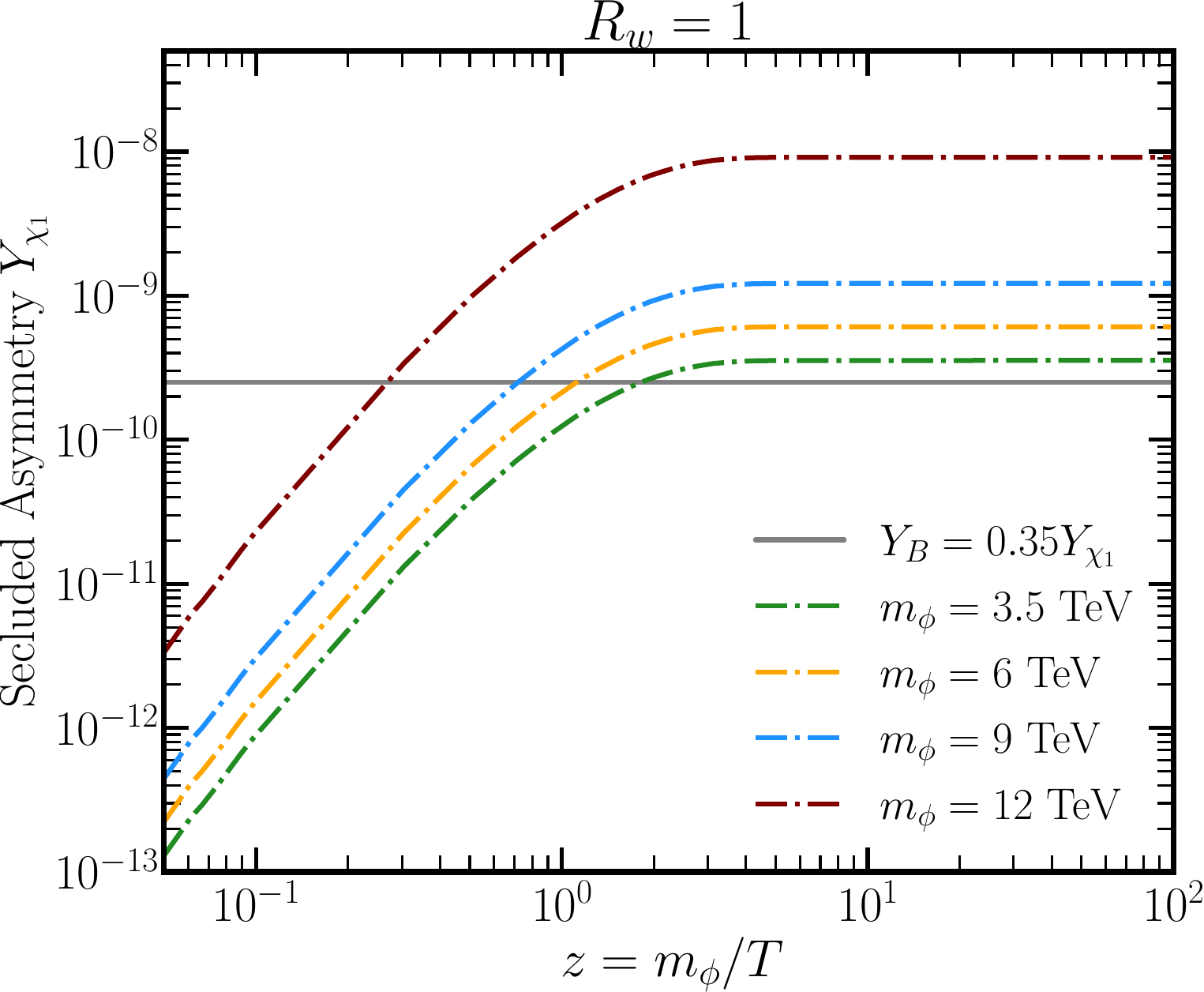} 
		\caption{\label{fig:Ychi1}The secluded $\chi_1$ asymmetry  with    the maximally resonant enhancement:  $\mathcal{S}_{\rm CP}(j=2, i=1)$. We   used a simple numerical example  $\text{Im}[y'_{11}y'^{*}_{21}y'^{*}_{12}y'_{22}]= -|y'_{11} y'_{21} y'_{12}y'_{22}|$, where $|y'_{11}|= |y'_{21}|$ is fixed by $R_w=1$ while $|y'_{12}|=|y'_{22}|=0.05$ is  fixed to solve Eq.~\eqref{eq:fphi} and suppress the NP corrections to lepton thermal masses.  The horizontal line corresponds to the observed BAU  that would be created via the secluded flavor effect. }
	\end{figure}

	To  proceed with the numerical computation of Eq.~\eqref{eq:KB}, we will work in    the weak washout regime,  where 
	 $\mathcal{W}$ can be neglected.  This regime can be characterized by the ratio  $R_w\equiv\Gamma_{\phi\to \ell+\chi_1}/H$  at $T=m_\phi$. Numerically, we apply~\cite{Buchmuller:2004nz}
	\begin{align}\label{eq:ratio-R}
		R_w= \frac{y_{1}^{\prime 2} m^*_{\rm Pl}}{16\pi m_\phi}\lesssim 1\,,
	\end{align}
	where $y_{j}^{\prime 2}\equiv \sum_i |y'_{ij}|^2$,  and $m^*_{\rm Pl}\approx 7.11\times 10^{17}$~GeV is a reduced Planck mass from the Hubble parameter $H$. 
	In this weak  washout  regime, Eq.~\eqref{eq:KB} reduces to 
	\begin{align}\label{eq:Y_chi1}
		\frac{\partial  Y_{\chi_1}}{\partial T}=-\frac{\mathcal{S}_{\rm CP}}{sHT}\,,
	\end{align}
	after the replacement $d/dt\to \partial/\partial t+3H$. The source term  depends on the scalar distribution function $f_\phi$ that should be  solved first.  
	To this aim, we will consider a large Yukawa coupling  $y'_2$  to simplify the scalar evolution via    $\phi \leftrightharpoons \ell +\chi_2$, which indicates that   decay/inverse decay will be stronger  than  gauge-scalar scattering at the leptogenesis epoch.   The leading scalar KB    equation  gives
	\begin{align}\label{eq:fphi}
		\frac{df_\phi}{dt}	=-\delta f_\phi	\frac{|y'_{2}|^2 m_\phi^2T}{8\pi E_\phi p_\phi}\ln\left(\frac{\cosh p_1}{\cosh p_2}\right)\,,
	\end{align}
	where $d/dt= \partial/\partial t-H p_\phi \partial/\partial p_\phi$, $p_{1,2}=(E_\phi\pm p_\phi)/(4T)$  and $p_\phi\equiv |\vec p_\phi|$. 
	Substituting the solution  $f_\phi$ into  $\mathcal{S}_{\rm CP}$ with an initial condition $f_\phi=f_\phi^{\rm eq}$ set at $z_0\equiv m_\phi/T_0=10^{-5}$, we  show in  Fig.~\ref{fig:Ychi1} the   evolution of  $Y_{\chi_1}$ with  the  maximal  enhancement from the muon flavor, i.e., $\mathcal{S}_{\rm CP}(j=2, i=1)$  and  a  maximal CP-violating Yukawa combination: $\text{Im}[y'_{11}y'^{*}_{21}y'^{*}_{12}y'_{22}]= -|y'_{11} y'_{21} y'_{12}y'_{22}|$.  
	
We see from Fig.~\ref{fig:Ychi1} that the  scalar  at  the TeV scale  is feasible to generate the baryon asymmetry via the secluded flavor effect [Eq.~\eqref{eq:YB}]. Fixing $R_w=1$, a heavier scalar $\phi$ will enhance the prefactor of Eq.~\eqref{eq:nonthermal-chi1-rate_1} and hence the secluded asymmetry $Y_{\chi_1}$. For   $m_\phi$ lower than the TeV scale, $Y_{\chi_1}$ can still be  enhanced  if  $R_w$ is larger. 
	 For  $R_w\gg 1$, however, leptogenesis will enter   the strong washout regime, where $f_{\chi_1}, f_{\bar \chi_1}$ and $\mathcal{W}$ cannot be neglected. In this regime,  we     expect that   the resonant forbidden CP asymmetry   still works with a larger Yukawa coupling $y'_1$, depending on the competition between $\mathcal{W}$ and $\mathcal{S}_{\rm CP}$.

	\textit{\textbf{Working regimes.}} 	It is interesting to note that Eq.~\eqref{eq:nonthermal-chi1-rate_1} is not singular under the hypothetical limit $y_j\to y_i$, as the NP contributions to $\tilde{m}^2$ come into play, which were not included in deriving Eq.~\eqref{eq:nonthermal-chi1-rate_1}. Including the NP corrections  and how they affect the final $Y_{\chi_1}$ would be  flavor dependent. For simplicity, we   fixed $|y'_{12}|=|y'_{22}|=0.05$ in Fig.~\ref{fig:Ychi1} to solve the scalar Boltzmann equation. This setup also serves to cancel the leading-order NP contributions in $\tilde{m}_2^2-\tilde{m}_1^2$, leaving $y_\mu^2|y'_{12}|^2$ as the next-to-leading-order correction. In more general situations, we find that the NP contributions can be neglected if $|y'_{22}|^2-|y'_{12}|^2\lesssim 10^{-6}$ holds. 
	Note that the  $\phi$-$\chi_1$ contribution  is typically negligible in $\tilde{m}_2^2-\tilde{m}_1^2$ since $\chi_1$ must be nonthermal in   scalar decay.   
	If the NP contributions are  larger than  the $y_i$ term,  we can infer from Eq.~\eqref{eq:nonthermal-chi1-rate_1}   that   $Y_{\chi_1}\propto\mathcal{O}(y'^2)$ instead of the   conventional situation  $Y_{\chi_1}\propto\mathcal{O}(y'^4)$. This is a new possibility, however, the precise computation requires solving the nontrivial time evolution of $\tilde{m}^2$. For an estimate, we take the Boltzmann distribution for $f_\phi$ at $m_\phi/T>1$ to determine the NP-corrected $\tilde{m}^2$~\cite{Beneke:2010dz,Kanemura:2024fbw}, and find that $Y_{\chi_1}=\mathcal{O}(10^{-10})$ can still be generated for $|y'_{12}|\simeq |y'_{22}|=\mathcal{O}(0.01), |y'_{22}|^2-|y'_{12}|^2\simeq10^{-4}$ if $R_w$ or $m_\phi$ is increased accordingly. 
	
The exemplary scalar   being in thermal equilibrium can still  work to generate a nonzero forbidden CP asymmetry if the nonthermal condition comes from $\chi_2$~\cite{SM}. In this case, the $\chi_2$-related Yukawa couplings are much smaller and  the NP corrections to lepton thermal masses become negligible. Scalar decay shown in this Letter  can also be switched to $\chi_1$ decay if $m_{\chi_1}>m_\phi$.   Moreover, unlike the traditional lepton flavor effects    occurring  far above the electroweak scale~\cite{Garbrecht:2012pq}, the resonant enhancement from Eq.~\eqref{eq:nonthermal-chi1-rate_1} applies even if   all  the  lepton flavors are in quasi-thermal equilibrium.  Therefore,   the resonant forbidden CP asymmetry features     low-scale leptogenesis, 
  making  production of $\phi, \chi$ attainable from low-energy experiments such as LHC and future Higgs factories. 
	The present example also  demonstrates  that  mixing from scalars is not necessary to induce CP asymmetry in  scalar decay,  allowing minimal   model buildings,  \textit{e.g.},  in   Dirac leptogenesis~\cite{Dick:1999je,Murayama:2002je} and seesaw type-II leptogenesis~\cite{Ma:1998dx}.  
	
	The main  purpose of  this Letter  is to exploit the maximal role of resummed leptons in generating CP asymmetries, and  the mechanism works for a wide class of particles $\phi,\chi$.  For example,   the particle set $(\phi,\chi)$   may be   identified as (scalar triplet, SM  leptons), (scalar doublet, sterile neutrinos), or  dark matter candidates that feature a novel  cogenesis mechanism for dark matter and the BAU.  
	
	When $(\phi,\chi)$ is identified as the SM Higgs and  sterile neutrinos,   it reduces to the prototype of the canonical type-I seesaw    leptogenesis. The contribution from the resonant  forbidden CP asymmetry may be noticeable in low-scale type-I leptogenesis, but the  $\chi_1$ asymmetry computed here cannot be directly  applied if the lepton number is  broken in the leptogenesis epoch.
	Nevertheless, the resonantly enhanced $\chi_1$ asymmetry   is valid  if  the   lepton-number breaking effect can be neglected, which is the case   for  pseudo-Dirac neutrinos~\cite{Wyler:1982dd} in low-scale seesaw frameworks~\cite{Mohapatra:1986bd,Branco:1988ex,Shaposhnikov:2006nn,Kersten:2007vk,Abada:2017ieq}, or for  relativistic  Majorana neutrinos~\cite{Akhmedov:1998qx,Abada:2018oly}.
	 Particularly,  the BAU can be generated by SM Higgs decay to GeV-scale Majorana neutrinos. It was pointed out in this case that a minimal   neutrino mass degeneracy at level of $10^{-5}$ is needed to enhance the CP asymmetry~\cite{Hambye:2016sby}. However, the thermal mass degeneracy from resummed leptons implies that an enhancement larger than $10^{5}$ is feasible  without neutrino mass degeneracy.  Note that for SM Higgs decay, the Higgs thermal   mass  cannot be neglected~\cite{Pilaftsis:1997jf}. For  massive $\phi$  before gauge symmetry breaking, however,  the  thermal mass at $\mathcal{O}(0.1) T$ from gauge interactions is not significant for low-scale leptogenesis culminating at $T\lesssim  m_\phi$ unless the couplings  in the scalar potential are  large~\cite{Cline:1995dg}. 
	 
For lepton-number  nonconservation,  one should build upon  mixed KB kinetic equations. Such a technique was considered in Ref.~\cite{Garbrecht:2012pq}, where    lepton flavor effects occurring  above $T=10^9$~GeV  could lead to a resonant enhancement at $\mathcal{O}(10^5)$.  However, the enhancement is dominated by gauge interactions rather than from  quadratic lepton Yukawa couplings, since the   lepton flavor effects     originate from  the interplay between  gauge and lepton Yukawa interactions. As a qualitative check, we have  confirmed that   the   mixed KB equation~\cite{Beneke:2010dz} with quasi-thermal leptons can lead to the consistent $1/(y^{2}+g^2 y^2)$  enhancement found here,  though the power counting of gauge and lepton Yukawa couplings is not so straightforward as the present treatment.  The development   with   the more general  mixed KB equations  goes beyond the current simple  approach   but deserves full consideration.  
	
\textit{\textbf{Conclusion}}. SM leptons at high temperatures can already give  rise to  a  large  plasma-induced CP asymmetry, which is an irreducible contribution to  low-scale leptogenesis.  The resonant forbidden CP asymmetry also provides  a new mechanism for leptogenesis   where  vacuum cuts or vacuum mass degeneracy are absent in the NP sector,  allowing minimal and testable model exploitation.

		\textit{\textbf{Acknowledgements.}}---
	We would like to thank Yushi Mura for frequent discussions during this project and Ke-Pan Xie for useful comments. 
	This project is supported by JSPS  Grant-in-Aid for JSPS Research Fellows No.~24KF0060. S.~K. is also supported in part by Grants-in-Aid for Scientific Research (KAKENHI) No.~23K17691 and No.~20H00160.
	
	\bibliographystyle{JHEP}
	\bibliography{Refs}
	
\onecolumngrid

\pagebreak

\setcounter{page}{1}

\begin{center}
	{\Large\textbf{Supplemental Material}}
\end{center}
\begin{center}
	{\large\textit{Resonant Forbidden CP Asymmetry from  Soft Leptons}}
\end{center}
\begin{center}
	{\large 	Shinya Kanemura  and Shao-Ping Li}
\end{center}

\section{CTP formalism and HTL resummation}\label{sec:CTP-HTL}
In the Closed-Timed-Path (CTP) formalism,  free propagators   in a spatially homogeneous and close-to-equilibrium plasma can be formulated by
\begin{align}\label{eq:S<}
	i\slashed{S}^{+-}(p)&\equiv i \slashed{S}^<(p)=-2\pi \delta(p^2-m^2)(\slashed{p}+m)\left[\theta(p_0)f(p_0)-\theta(-p_0)(1-\bar f(-p_0))\right],
	\\[0.2cm]
	i\slashed{S}^{-+}(p)&\equiv	i \slashed{S}^>(p)=-2\pi \delta(p^2-m^2)(\slashed{p}+m)\left[-\theta(p_0)(1-f(p_0)+\theta(-p_0)\bar f(-p_0)\right],\label{eq:S>}
	\\[0.2cm]
	i\slashed{S}^{++}(p)&\equiv i\slashed{S}^T(p)=\frac{i(\slashed{p}+m)}{p^2-m^2+i\epsilon}-2\pi \delta(p^2-m^2)(\slashed{p}+m)\left[\theta(p_0)f(p_0)+\theta(-p_0)\bar f(-p_0)\right],\label{eq:ST}
	\\[0.2cm]
	i\slashed{S}^{--}(p)&\equiv	i\slashed{S}^{\bar T}(p)=-\frac{i(\slashed{p}+m)}{p^2-m^2-i\epsilon}-2\pi \delta(p^2-m^2)(\slashed{p}+m)\left[\theta(p_0)f(p_0)+\theta(-p_0)\bar f(-p_0)\right],\label{eq:STbar}
\end{align}
for  massive Dirac  fermions, and
\begin{align}\label{eq:freeProp-G}
	iG^{+-}(p)&\equiv	i G^<(p)=2\pi \delta(p^2-m^2)\left[\theta(p_0)f(p_0)+\theta(-p_0)(1+\bar f(-p_0))\right],
	\\[0.2cm]
	iG^{-+}(p)&\equiv	i G^>(p)=2\pi \delta(p^2-m^2)\left[\theta(p_0)(1+f(p_0))+\theta(-p_0)\bar f(-p_0)\right],
	\\[0.2cm]
	iG^{++}(p)&\equiv	iG^T(p)=\frac{i}{p^2-m^2+i\epsilon}+2\pi \delta(p^2-m^2)\left[\theta(p_0)f(p_0)+\theta(-p_0)\bar f(-p_0)\right],
	\\[0.2cm]
	iG^{--}(p)&\equiv	iG^{\bar T}(p)=-\frac{i}{p^2-m^2-i\epsilon}+2\pi \delta(p^2-m^2)\left[\theta(p_0)f(p_0)+\theta(-p_0)\bar f(-p_0)\right],
\end{align}
for massive  scalars.  Here, $\pm$ denote the  thermal indices with  $\lessgtr$   representing the Wightman functions and $T(\bar T)$ the (anti) time-ordered propagators, and  $f,\bar f$  are the distribution functions of particles and antiparticles~\cite{Prokopec:2003pj,Prokopec:2004ic}.  
In thermal equilibrium,  we have
\begin{align}
	f(p_0)=\bar f(p_0)=\frac{1}{e^{p_0/T}\pm 1}\,
\end{align}
for fermions ($+$) and bosons ($-$). Besides, the Kubo–Martin–Schwinger (KMS)  relations hold $\slashed{S}^>(p)=-e^{p_0/T}\slashed{S}^<, G^>(p)=e^{p_0/T}G^<(p)$.    

The resummed scalar propagator concerned in the Letter  can   be described by adding the    thermal mass effect to the vacuum mass, which is nevertheless subdominant.  For SM singlets $\chi_{1,2}$, the thermal mass effects arise from the small Yukawa couplings $y'$ and we also treat them   as subdominant thermal corrections to vacuum masses. 
To derive the resummed lepton propagators,  we use the approximation that leptons are in   thermal equilibrium during the leptogenesis epoch.  The resummed Wightman functions are written as\footnote{We will use the same symbol for free and resummed propagators.}
\begin{align}\label{eq:tilde-S<>}
	\slashed{S}^<(p)&=-f(p_0)\left(\slashed{S}^R-\slashed{S}^A\right),
	\\[0.2cm]
	\slashed{{S}}^>(p)&=\left[1-f(p_0)\right]\left(\slashed{{S}}^R-\slashed{{S}}^A\right),
\end{align}
where $\slashed{{S}}^R$ and $\slashed{{S}}^A$ are the resummed  retarded and advanced propagators satisfying   
\begin{align}\label{eq:S-relations}
	\slashed{{S}}^T- \slashed{{S}}^{\bar T}=\slashed{{S}}^R+\slashed{{S}}^A, \quad \slashed{{S}}^>- \slashed{{S}}^<=\slashed{{S}}^R-\slashed{{S}}^A\,.
\end{align}
The resummed retarded propagator is generally  written as~\cite{Weldon:1982bn,Li:2023ewv}
\begin{align}\label{eq:tildeS^R}
	\slashed{{S}}^{R}(p)	&=P_{L}\frac{(1+a_{})\slashed{p}+b_{}\slashed{u}}{\left[(1+a_{})p_0+b_{}\right]^{2}-\left[(1+a_{})|\vec{p}|\right]^{2}}P_{R}
	\\[0.2cm]
	&\equiv \sum_{\pm}\frac{1}{\text{Re}\Delta_{\pm}+i \text{Im}\Delta_{\pm}}{P}_\pm\,,\label{eq:tildeS^R-def}
\end{align}
where $u_{\mu}$ is the four-velocity of the plasma normalized by $u_{\mu}u^{\mu}=1$  with $u_{\mu}=(1,0,0,0)$ in the rest frame. Note that we have omitted the $i\text{sign}(p_0)\epsilon$ prescription for resummed propagators.  The modified dispersion relation and thermal width are determined by the real and imaginary parts, respectively:
\begin{align}
	\text{Re}\Delta_{\pm}(p)&\equiv(1+\text{Re} a_{})(p_{0}\pm |\vec{p}|)+\text{Re}b_{}\,,
	\\[0.2cm]
	\text{Im}\Delta_{\pm}(p)&\equiv \text{Im} a_{}(p_{0}\pm |\vec{p}|)+\text{Im}b_{}\,,
\end{align}
and  $P_\pm$ denotes the decomposition of helicity eigenstates~\cite{Braaten:1990wp}
\begin{align}
	P_\pm\equiv P_{L}\frac{\gamma^{0}\pm \vec{e}_{p}\cdot\vec{\gamma}}{2}P_{R}\, ,
\end{align}
with $\vec{e}_p\equiv\vec{p}/|\vec{p}|$. 
The resummed advanced propagator can be obtained via $a,b\to a^*, b^*$,  giving rise to 
\begin{align}\label{eq:SR-SA}
	\slashed{{S}}^{R}-\slashed{{S}}^{A}	=\sum_{\pm}\frac{-2i\text{Im}\Delta_{\pm}}{(\text{Re}\Delta_{\pm})^{2}+(\text{Im}\Delta_{\pm})^{2}}P_\pm\, .
\end{align}
The real coefficients $\text{Re}a_{}, \text{Re}b_{}$ are well known in the HTL approximation~\cite{Weldon:1982bn,Li:2023ewv} 
\begin{align}\label{eq:aRi}
	\text{Re}a_{i}&=\frac{\tilde{m}^2_{i}}{|\vec{p}|^2}\left[1+\frac{p_0}{2|\vec{p}|}\ln\left(\frac{p_0-|\vec{p}|}{p_0+|\vec{p}|}\right)\right],
	\\[0.3cm]
	\text{Re}b_{i}&=-\frac{\tilde{m}^2_{i}}{|\vec{p}|}\left[\frac{p_0}{|\vec{p}|}-\frac{1}{2}\left(1-\frac{p_0^2}{|\vec{p}|^2}\right)\ln\left(\frac{p_0-|\vec{p}|}{p_0+|\vec{p}|}\right)\right]\,,\label{eq:bRi}
\end{align}
with the properties $\text{Re}a_{i}(p_0)=\text{Re}a_{i}(-p_0), \text{Re}b_{i}(p_0)=-\text{Re}b_{i}(-p_0)$, and the 
thermal masses of leptons:
\begin{align}\label{eq:thermalmass}
	\tilde{m}_{i}^2=\left(\frac{3}{32}g_2^2+\frac{1}{32}g_1^2+\frac{1}{16}y_{i}^2\right)T^2\,.
\end{align}

\begin{figure}[t]
	\centering
	\includegraphics[scale=0.7]{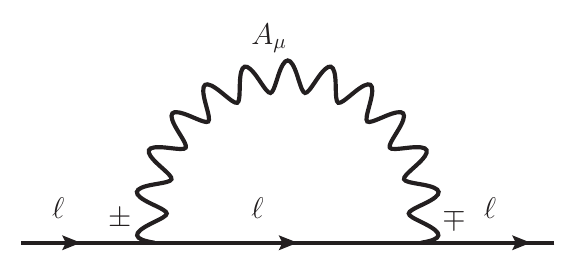} 
	\caption{\label{fig:gauge-1PI} Contributions to thermal widths of leptons from gauge interactions, where $A_\mu=W_\mu, B_\mu$ denote gauge $SU(2)_L$ and $U(1)_Y$ bosons and $\pm$  the thermal indices.  }
\end{figure}

In the zero-width approximation,  $\text{Im}\Delta_\pm\to 0$, the pole of both $\slashed{{S}}^R$ and $\slashed{{S}}^R-\slashed{{S}}^A$ (hence $\slashed{{S}}^{<,>}$) is determined by $\text{Re}\Delta_{\pm}=0$, which gives
\begin{align}\label{eq:pole4}
	p_0= \mp \left[|\vec p|+\frac{\tilde{m}^2}{|\vec p|}-\frac{\tilde{m}^4}{2|\vec p|^3}\log\left(\frac{2|\vec p|^2}{\tilde{m}^2}\right)+\mathcal{O}(\tilde{m}^6)\right].
\end{align}
At $\mathcal{O}(\tilde{m}^2)$, the dispersion relations can be well approximated by two modes: $p_0^2-|\vec p|^2\approx 0$ and  $p_0^2-|\vec{p}|^2\approx 2\tilde{m}^2$~\cite{Kiessig:2010pr,Drewes:2013iaa,Li:2023ewv}.  The first mode is flavor blind and  will not give rise to the  forbidden CP asymmetry that is lepton-flavor dependent.
Neglecting $\mathcal{O}(g^4)$ contributions to the pole, we would  arrive at the onshell propagation from the second mode:
\begin{align}\label{eq:tildeS^R-onshell}
	i\slashed{{S}}^{R}_{i}(p)\Big|_{\rm onshell}&\approx \pi\text{sign}(p_{0}) \delta(p^2-2\tilde{m}_i^2)P_L\slashed{p}P_R\,,
	\\[0.2cm]
	i\slashed{{S}}_i^{<}(p)\Big|_{\rm onshell}&\approx -2\pi\text{sign}(p_0) f(p_0)\delta(p^2-2\tilde{m}_i^2)P_L\slashed{p}P_R\,,\label{eq:tildeS^<-onshell}
	\\[0.2cm]
	i\slashed{{S}}_i^{>}(p)\Big|_{\rm onshell}&\approx 2\pi\text{sign}(p_0)\left[1-f(p_0)\right]\delta(p^2-2\tilde{m}_i^2)P_L\slashed{p}P_R\,,\label{eq:tildeS^>-onshell}
\end{align}
where  the resummed Wightman functions have the same form as in the free case.  

To see the  consequence of including $\text{Im}\Delta_\pm$ at higher orders of  gauge couplings, which is important to check the stability of the resonant forbidden CP asymmetry, we next go beyond the zero-width approximation by including the finite width from $\text{Im}\Delta_\pm$ perturbatively.  To this aim, we   first calculate the imaginary coefficients $\text{Im}a_{},\text{Im}b_{}$.  Consider the leading-order contribution from the gauge 1PI diagrams shown in Fig.~\ref{fig:gauge-1PI}. The imaginary part of the  retarded amplitude for the SM leptons can be calculated via
\begin{align}
	\text{Im}\Sigma^R=-\frac{i}{2}\left(\Sigma^>-\Sigma^<\right)\,,
\end{align}
and the imaginary coefficients $\text{Im}a_{},\text{Im}b_{}$ read
\begin{align}\label{eq:ImabR}
	\text{Im}a_{}	&=\frac{1}{2|\vec{p}|^{2}}\left(\text{Tr}[\slashed{p}\text{Im}\Sigma^{R}(p)]-p_{0}\text{Tr}[\slashed{u}\text{Im}\Sigma^{R}(p)]\right),
	\\[0.2cm]
	\text{Im}	b_{}	&=-\frac{1}{2|\vec{p}|^{2}}\left(p_{0}\text{Tr}[\slashed{p}\text{Im}\Sigma^{R}(p)]-p^{2}\text{Tr}[\slashed{u}\text{Im}\Sigma^{R}(p)]\right).
\end{align}
A straightforward calculation of Fig.~\ref{fig:gauge-1PI}   at $\text{Re}\Delta_\pm=0$  gives 
\begin{align}
	\text{Im}\Delta_\pm= \kappa g^2 \frac{\tilde{m}^2}{|\vec{p}|}\,,
\end{align}
where  $g^2\equiv 3g_2^2+g_1^2$ includes the contributions from $SU(2)_L$ and $U(1)_Y$ gauge bosons,  and  for simplicity we have taken $T=|\vec p|$ to turn the loop integration into a numerical coefficient $\kappa\approx 4\times 10^{-3}$.  
We see that the leading dependence  of $\text{Im}\Delta_\pm$ on gauge couplings is at $\mathcal{O}(g^4)$, which is indeed a higher-order effect for Eqs.~\eqref{eq:tildeS^R-onshell}-\eqref{eq:tildeS^>-onshell}.

As will  be shown in Sec.~\ref{sec:CR},  the product ${S}_i^R\times {S}^{<}_j$ contributes to the resonance.  Besides, the kinetics  shown in Sec.~\ref{sec:CR}  selects  one of the poles from Eq.~\eqref{eq:pole4}, so   only the $ P_+\times  P_+$ or $P_-\times  P_-$ component in the product ${S}_i^R\times {S}^{<}_j$ will contribute to the resonance.  Here, the unslashed propagators denote the c-numbers  from the slashed propagators, e.g., the $P_{+}$ component of $S^R$ is defined as: $[S^R]_+\equiv 1/(\text{Re}\Delta_+ +i \text{Im}\Delta_+)$.
Let us consider the $ P_+\times  P_+$ component at $\mathcal{O}(g^2)$, though  the analysis for  the $ P_-\times  P_-$ component is essentially the same. Using Eqs.~\eqref{eq:tildeS^<-onshell} and~\eqref{eq:tildeS^R-def},  and factoring out the distribution functions, we arrive at 
\begin{align}\label{eq:g2product++}
	\left[S_i^R(p)\times S^{<}_j(p)\right]_{++}= 2\pi i\frac{\delta(\text{Re}\Delta_{+j})}{\text{Re}\Delta_{+i}}=2\pi i\frac{|\vec p|}{\tilde{m}_i^2-\tilde{m}_j^2}\delta(p_0+\omega_j)\,,
\end{align}
where $\omega_j=(|\vec p|^2+2\tilde{m}_j^2)^{1/2}$ corresponds to    the  pole given in Eq.~\eqref{eq:pole4} up to  $\mathcal{O}(g^2)$.   Equation~\eqref{eq:g2product++} demonstrates the origin of the  resonant enhancement at $\mathcal{O}(g^2)$. 

When including $\text{Im}\Delta_\pm$ at $\mathcal{O}(g^4)$, we can analyze the resonance by using the  Cauchy's residue theorem.   The $P_+\times P_+$ component  of the $S_i^R\times S^{<}_j$ product now has a pole  from $[S^{<}_j]_+$:
\begin{align}\label{eq:pole-4}
	p_0= - \left[ |\vec p|+\frac{\tilde{m}_j^2}{p}-\frac{\tilde{m}_j^4}{2|\vec p|^3}\log\left(\frac{2|\vec p|^2}{\tilde{m}_j^2}\right)+i\kappa g^2\frac{\tilde{m}_j^2}{|\vec p|}\right].
\end{align}
Choosing the lower half-plane contour that encloses the above pole would give rise to 
\begin{align}\label{eq:g4product++}
	\left[{S}_i^R(p)\times {S}^{<}_j(p)\right]_{++}\xrightarrow{\oint_c} 2\pi i\frac{|\vec p|}{\tilde{m}_i^2-\tilde{m}_j^2+i\kappa g^2(\tilde{m}_i^2-\tilde{m}_j^2)}\,.
\end{align}
Here $\xrightarrow{\oint_c}$ indicates that the right-hand side is the result that would be obtained after performing the contour integration,  with a factor of $-2\pi i$ from the residue theorem. We see that the next-to-leading-order contribution beyond $\tilde{m}_i^2-\tilde{m}_j^2$ is  $\kappa g^2(\tilde{m}_i^2-\tilde{m}_j^2)$, which is at $\mathcal{O}(g^2 y^2)$.  Given  $\kappa g^2(\tilde{m}_i^2-\tilde{m}_j^2)\ll \tilde{m}_i^2-\tilde{m}_j^2$, the result shown in  Eq.~\eqref{eq:g4product++}   will approximately reduce to that in Eq.~\eqref{eq:g2product++} after the full momentum integration. Therefore, including the $\mathcal{O}(g^{4})$ contribution both in the dispersion relation and thermal width will not destroy the resonance, since the dependence on the flavor-universal gauge couplings is canceled.   In fact, going to $\mathcal{O}(g^{n>4})$, one can  repeat the above perturbative analysis, and the gauge-coupling dependence will be canceled in the denominator   order by order, leaving terms involving the mixed couplings $g^a y^b$ with $a+b=n$.

Note that there is another  pole   in the $P_+\times  P_+$ component  of  ${S}_i^R\times {S}^{<}_j$, which comes from $[S^R_i]_+$.  However, the contribution from this pole does not exhibit a resonance enhanced by $\mathcal{O}(y^{-2})$.  One   can check this by   applying the residue theorem with  the pole of $[S^R_i]_+$, which is given by Eq.~\eqref{eq:pole-4} with the replacement $\tilde{m}_j\to \tilde{m}_i$. 

The above derivation shows that we should take the thermal widths of lepton flavors in a way consistent with the HTL perturbative expansion. Using the zero-width approximation  for the Wightman functions   but inserting the thermal width in the retarded propagator is not a  consistent  treatment, since the former is not valid  when keeping the  $\mathcal{O}(g^4)$ terms. The identical dependence on   gauge couplings  of different lepton flavors ensures that the resonance from $1/(\tilde{m}_i^2-\tilde{m}_j^2)$ is stable under    higher-order gauge   corrections to lepton dispersion relations and thermal widths.

\section{Kadanoff-Baym  equations}
The KB equation used in the Letter is derived from the full  kinetic  equation, which 
in  Wigner space is given by~\cite{Prokopec:2003pj}
\begin{align}
	(\slashed{p}+\frac{i}{2}\slashed{\partial}_x-m_{\chi_1} e^{-i \overleftarrow{\partial_x}\cdot \partial_p /2})\slashed{S}_{\chi_1}^{<,>}-e^{-i \diamond/2}\{\slashed\Sigma_h\}\{\slashed{S}_{\chi_1}^{<,>}\}-e^{-i \diamond/2}\{\slashed\Sigma_{\chi_1}^{<,>}\}\{\slashed S_h\}=\mathcal{C}_{\chi_1}\,,
\end{align}
where $\partial_x\equiv \partial/\partial x, \partial_p\equiv \partial/\partial p$ are the spacetime and 4-momentum derivatives, and  the symbol $\diamond$ denotes the so-called diamond operator with $\diamond\{A\}\{B\}=(\partial_x A) (\partial_p B)-(\partial_p A) (\partial_x B)$. The propagator $\slashed S_h$ and the self-energy amplitude $\slashed\Sigma_h$ can be written in terms of time and anti-time ordered quantities:
\begin{align}
	\slashed S_h\equiv \frac{1}{2}(\slashed S_{\chi_1}^T-\slashed S_{\chi_1}^{\bar T})\,,\quad
	\slashed\Sigma_h\equiv \frac{1}{2}(\slashed\Sigma_{\chi_1}^T-\slashed\Sigma_{\chi_1}^{\bar T})\,. 
\end{align} 
$\mathcal{C}_{\chi_1}$ is the collision term  accounting for gain and loss rates of $\chi_1$:
\begin{align}
	\mathcal{C}_{\chi_1}=\frac{1}{2}e^{-i \diamond/2}\left(\{\slashed\Sigma_{\chi_1}^>\}\{\slashed S_{\chi_1}^<\}-\{\slashed \Sigma_{\chi_1}^<\}\{ S_{\chi_1}^>\}\right).
\end{align}
At zeroth order of gradient expansion: $e^{-i\diamond/2}\approx 1$,  the kinetic equation with a constant mass reduces to
\begin{align}
	(\slashed{p}+\frac{i}{2}\slashed{\partial}_x-\tilde{m}_{\chi_1}) \slashed S_{\chi_1}^{<,>}-\slashed\Sigma_{\chi_1}^{<,>}\times \slashed S_h=\frac{1}{2}\left(\slashed \Sigma_{\chi_1}^>\times  \slashed S_{\chi_1}^<-\slashed\Sigma_{\chi_1}^<\times \slashed S_{\chi_1}^>\right),
\end{align}
where $\tilde{m}_{\chi_1}\equiv m_{\chi_1}+ \slashed\Sigma_h$ includes vacuum and finite-temperature self-energy corrections to the dispersion relation of $\chi_1$.   In the KB ansatz followed by  the onshell quasiparticle approximation, we have  $(\slashed{p}-\tilde{m}_{\chi_1})\slashed S_{\chi_1}^{<,>}=\slashed\Sigma_{\chi_1}^{<,>}\times \slashed S_h\approx 0$. Therefore, the KB kinetic equation reduces to
\begin{align}\label{eq:0th-KB}
	\gamma^0 \frac{d}{d t} (i\slashed S_{\chi_1}^{<,>})=(-i\slashed\Sigma_{\chi_1}^{>})(i\slashed S_{\chi_1}^{<})-(-i\slashed \Sigma_{\chi_1}^{<})(i\slashed S_{\chi_1}^>)\,,
\end{align}
where $-i$ attached to $\slashed{\Sigma}^{<,>}$   highlights the convention for   self-energy amplitudes.
Using Eqs.~\eqref{eq:S<}-\eqref{eq:S>} for $\slashed S_{\chi_1}^{<,>}$, taking the Dirac trace on both sides of the above equation, and  integrating over $d^4p_{\chi_1}/(2\pi)^4$,  we will arrive at the KB equation of $\chi_1$ used in the Letter.

Note that the  KB ansatz with the quasiparticle approximation is consistent with  the pole equation~\cite{Prokopec:2003pj,Berges:2004yj}
\begin{align}\label{eq:constraint}
	(\slashed{p}+\frac{i}{2}\slashed{\partial}_x-m_{\chi_1}e^{-i \overleftarrow{\partial_x}\cdot \partial_p /2}) \slashed S_{\chi_1}^{r,a}-e^{-i \diamond/2}\{\slashed\Sigma_{\chi_1}^{r,a}\}\{\slashed S_{\chi_1}^{r,a}\}=\bf{1}\,,
\end{align}
where 
\begin{align}
	\slashed S_{\chi_1}^{r,a}&=\frac{1}{2}\left[\slashed S_{\chi_1}^T-\slashed S_{\chi_1}^{\bar T}\pm \left(\slashed S_{\chi_1}^>-\slashed S_{\chi_1}^<\right)\right]=\slashed S_h\pm \frac{1}{2}\left(\slashed S_{\chi_1}^>-\slashed S_{\chi_1}^<\right),
	\\[0.2cm]
	\slashed\Sigma_{\chi_1}^{r,a}&=\frac{1}{2}\left[\slashed \Sigma_{\chi_1}^T-\slashed\Sigma_{\chi_1}^{\bar T}\pm \left(\slashed\Sigma_{\chi_1}^>-\slashed\Sigma^<\right)\right]=\slashed\Sigma_h \pm \frac{1}{2}\left(\slashed\Sigma_{\chi_1}^>-\slashed\Sigma_{\chi_1}^<\right).
\end{align}
To check this, we  take the difference   $\slashed S^r-\slashed S^a$ via Eq.~\eqref{eq:constraint}  in the  onshell limit: $\slashed S_h=0$, which at zeroth order of gradient expansion gives rise to 
\begin{align}
	(\slashed{p}+\frac{i}{2}\slashed{\partial}_x-\tilde{m}_{\chi_1})(\slashed S_{\chi_1}^>-\slashed S_{\chi_1}^<)=0\,.
\end{align}
Further using the onshell limit $(\slashed{p}-\tilde{m}_{\chi_1})\slashed S_{\chi_1}^{<,>}= 0$, we see that   $\slashed{\partial}_x \slashed S_{\chi_1}^>=\slashed{\partial}_x \slashed S_{\chi_1}^<$, which   is consistent with    Eq.~\eqref{eq:0th-KB}.

The full KB equation for scalar  $\phi$ reads~\cite{Prokopec:2003pj}
\begin{align}\label{eq:scalar-KB-full}
	e^{-i \diamond/2}\{p^2-\tilde{m}_\phi^2\}\{G_\phi^{<,>}\}-e^{-i \diamond/2}\{\Pi_\phi^{<,>}\}\{G_{h}\}=\frac{1}{2}e^{-i \diamond/2}\left(\{ \Pi_\phi^>\}\{G_\phi^<\}-\{\Pi_\phi^<\}\{G_\phi^>\}\right),
\end{align}
where $\tilde{m}_\phi^2 \equiv m_\phi^2+\Pi_{h}$ with
\begin{align}
	\Pi_{h}\equiv \frac{1}{2}(\Pi_\phi^T-\Pi_\phi^{\bar T})\,,  \quad G_{h}=\frac{1}{2}(G_\phi^T-G_\phi^{\bar T})\,.
\end{align}  
Similar to the fermion case, we have $(p^2-\tilde{m}_\phi^2)G_\phi^{<,>}=\Pi_\phi^{<,>}\times G_{h}\approx0$ in the onshell limit.  Then expanding the gradient  up to first order and neglecting the spacetime/4-momentum derivatives of loop amplitudes,  we arrive at  the Boltzmann equation given in the Letter, with the decay and inverse decay rates equivalent to what would be derived from the $S$-matrix formalism.  

Note that the onshell approximation  $(p^2-\tilde{m}_\phi^2)G_\phi^{<,>}\approx 0$ is equivalent to  $\Pi_\phi^{<,>}\times G_{h}\approx0$ at zeroth order. To see this, we can derive the Hermitian part of Eq.~\eqref{eq:scalar-KB-full}, 
\begin{align}\label{eq:scalar-KB-full-hc}
	e^{i \diamond/2}\{p^2-\tilde{m}_\phi^2\}\{G_\phi^{<,>}\}-e^{i \diamond/2}\{\Pi_\phi^{<,>}\}\{G_{h}\}=-\frac{1}{2}e^{i \diamond/2}\left(\{ \Pi_\phi^>\}\{G_\phi^<\}-\{\Pi_\phi^<\}\{G_\phi^>\}\right),
\end{align}
where we have used the Hermitian conditions $(iG_\phi^{<,>})^\dagger=iG_\phi^{<,>}$, $(i\Pi_\phi^{<,>})^\dagger=i\Pi_\phi^{<,>}$ and $G_h^\dagger=G_h$.   Then taking the sum   of Eq.~\eqref{eq:scalar-KB-full} and Eq.~\eqref{eq:scalar-KB-full-hc} and expanding to the first-order gradient, we arrive at the so-called constraint equation:
\begin{align}
	(p^2-\tilde{m}_\phi^2)G_\phi^{<,>}+\frac{i}{4}\diamond\left(\{\Pi_\phi^>\}\{G_{\phi}^<\}-\{\Pi_\phi^<\}\{G_\phi^>\}\right)=\Pi_\phi^{<,>}\times G_{h}\,.
\end{align}
Therefore,    the onshell limit  $(p^2-\tilde{m}_\phi^2)G_\phi^{<,>}=0$ indicates $\Pi_\phi^{<,>}\times G_{h}=0$ at zeroth order. Alternatively,  the dissipative term $\Pi_\phi^{<,>}\times G_{h}$ is of the same order as the   spacetime/4-momentum derivatives of loop amplitudes in the onshell limit, such that  the kinetic equation for the scalar distribution function will reduce to the Boltzmann equation.

\section{CP-violating source}\label{sec:CR}

The discussion in Sec.~\ref{sec:CTP-HTL} demonstrated that the thermal widths of resummed leptons are   smaller than the thermal mass effects near the quasiparticle pole, and the resonance due to the thermal mass difference is insensitive to    higher-order gauge coupling corrections. Given this, we will calculate the CP-violating source by neglecting contributions from $\mathcal{O}(g^4)$.

\begin{figure}[t]
	\centering
	\includegraphics[scale=0.9]{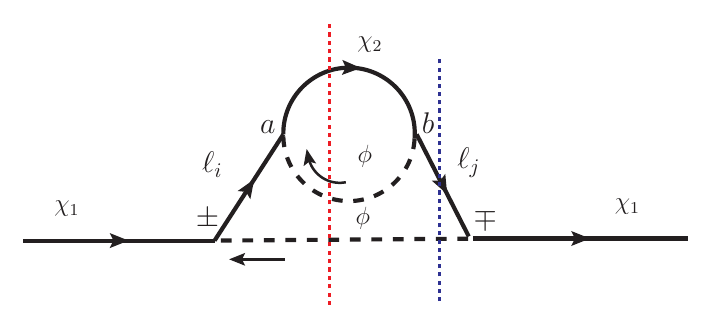} 
	\caption{\label{fig:lep-cuts} The two-loop self-energy diagram of $\chi_1$ contributing to the forbidden CP asymmetry.  The outer vertices are fixed (from left to right) by $+-$ and $-+$ while the inner vertices $a,b$ are summed over thermal indices $\pm$. The arrows for the scalar propagators denote the momentum flow, and   the fermion arrows align with the momentum flow.   Qualitatively, the red cutting line induces scattering processes mediated by leptons, and the blue cutting line corresponds to scalar decay and inverse decay if $m_\phi>m_{\chi_{1}}, m_{\chi_2}$. }
\end{figure}

The self-energy diagram of  $\chi_1$ that contributes to the forbidden CP asymmetry is shown in Fig.~\ref{fig:lep-cuts}, with the following amplitudes
\begin{align}\label{eq:S<_chi1}
	i\slashed{\Sigma}_{\chi_1}^<(p_{})&=2y'_{i1}y_{j1}^{\prime *}\int \frac{d^4 p_\ell}{(2\pi)^4}\frac{d^4 p_\phi}{(2\pi)^4}(2\pi)^4\delta^4(p_{}-p_\ell+p_\phi)P_L i\slashed{{S}}^<_{\ell_{ij}}P_R iG_\phi^>\,,
	\\[0.3cm]
	i\slashed{\Sigma}_{\chi_1}^>(p_{})&=2y'_{i1}y^{\prime *}_{j1}\int \frac{d^4 p_\ell}{(2\pi)^4}\frac{d^4 p_\phi}{(2\pi)^4}(2\pi)^4\delta^4(p_{}-p_\ell+p_\phi)P_L i\slashed{{S}}^>_{\ell_{ij}}P_R iG_\phi^<\,,\label{eq:S>_chi1}
\end{align}
where $p$ denotes the $\chi_1$ momentum. We have used the convention $-i\Sigma$ for the loop amplitudes, and the factor of 2 comes from the gauge $SU(2)_L$ degeneracy. It is worth mentioning that  since the KB equation is derived by the 2PI effective action, where
\begin{align}\label{eq:Self_from_2PI}
	i\slashed\Sigma_{}^{ab}=ab \frac{\delta \Gamma_2}{\delta \slashed S^{ba}}\,,
\end{align}   
the minus sign from the Feynman rule of  the  \textit{negative}  thermal vertex is compensated by the factor $ab$ from the above functional derivative. Therefore, when we use Feynman rules to  write down the amplitudes  $i\slashed\Sigma^{<,>}$ on the right-hand side of the KB equation,   there is no minus sign from the  \textit{negative}  thermal vertex, but amplitudes from the inner loop do carry such minus signs.  Under this convention,    the  Schwinger-Dyson equation  indicates
\begin{align}\label{eq:SD-relations}
	\slashed S^T+\slashed S^{\bar T}=\slashed S^<+\slashed S^>\,,\quad
	\slashed\Sigma^T+\slashed\Sigma^{\bar T}=\slashed\Sigma^<+\slashed\Sigma^>\,,
\end{align} 
which will be used to  simplify the collision rates in the KB equation. 

In Eqs.~\eqref{eq:S<_chi1}-\eqref{eq:S>_chi1}, $i\slashed{S}^<_{\ell_{ij}}$ denotes the resummed lepton propagators. Summing over  the thermal indices $\pm$ for $a,b$ in Fig.~\ref{fig:lep-cuts},  we can write
\begin{align}\label{eq:tildeS<Lij}
	i\slashed{{S}}^<_{\ell_{ij}}&=(i\slashed{{S}}^R_{\ell_i}) (-i\slashed{\Sigma}^T_{\ell_{ij}})(i\slashed{{S}}^<_{\ell_j})+(i\slashed{{S}}^R_{\ell_i})(-i\slashed{\Sigma}^<_{\ell_{ij}})(i\slashed{{S}}^R_{\ell_j})+(i\slashed{{S}}^<_{\ell_i})(-i\slashed{\Sigma}^<_{\ell_{ij}})(i\slashed{{S}}^R_{\ell_j})
	\nonumber\\[0.2cm]
	&-(i\slashed{{S}}^R_{\ell_i})(-i\slashed{\Sigma}^<_{\ell_{ij}})(i\slashed{{S}}^>_{\ell_j})-(i\slashed{{S}}^<_{\ell_i})(-i\slashed{\Sigma}^{\bar T}_{\ell_{ij}})(i\slashed{{S}}^R_{\ell_j})\,,
	\\[0.2cm]
	i\slashed{{S}}^>_{\ell_{ij}}&=(i\slashed{{S}}^>_{\ell_i})(-i\slashed{\Sigma}^T_{\ell_{ij}})(i\slashed{{S}}^R_{\ell_j})+(i\slashed{{S}}^R_{\ell_i})(-i\slashed{\Sigma}^>_{\ell_{ij}})(i\slashed{{S}}^R_{\ell_j})+(i\slashed{{S}}^R_{\ell_i})(-i\slashed{\Sigma}^>_{\ell_{ij}})(i\slashed{{S}}^<_{\ell_j})
	\nonumber\\[0.2cm]
	&-(i\slashed{{S}}^>_{\ell_i})(-i\slashed{\Sigma}^>_{\ell_{ij}})(i\slashed{{S}}^R_{\ell_j})-(i\slashed{{S}}^R_{\ell_i})(-i\slashed{\Sigma}^{\bar T}_{\ell_{ij}})(i\slashed{{S}}^>_{\ell_j})\,,\label{eq:tildeS>Lij}
\end{align}  
where  Eqs.~\eqref{eq:S-relations} and~\eqref{eq:SD-relations}  have been used and the minus sign   from thermal index $a$ or $b$  being of  the \textit{negative} type has been included.  In addition, we have applied Eqs.~\eqref{eq:tildeS^<-onshell}-\eqref{eq:tildeS^>-onshell} such that $i {{S}}^<_{\ell_i}(p)\times i {{S}}^>_{\ell_j}(p)\propto \delta(p^2-2\tilde m_i^2)\delta(p^2-2\tilde m_j^2)=0$.  
Given $\slashed{{S}}^{>}_{\ell_{ij}}\propto y_{i2}^{\prime *}y'_{j2}$, we play the trick of interchanging   dummy indices $i, j$, so that  Eq.~\eqref{eq:S>_chi1} becomes
\begin{align}
	i\slashed{\Sigma}_{\chi_1}^{>}(p_{})
	=2(y'_{i1}y^{\prime *}_{j1}y_{i2}^{\prime *}y'_{j2})^* \int   \frac{d^4 p_\ell}{(2\pi)^4}\frac{d^4 p_\phi}{(2\pi)^4}(2\pi)^4\delta^4(p_{}-p_\ell+p_\phi)P_L i\slashed{S}^>_{\ell_{ji}}P_R iG_\phi^<\,,
\end{align}
where  the Yukawa couplings are factored out from $-i\slashed{\Sigma}_{\ell_{ij}}\equiv y^{\prime *}_{i2}y'_{j2}\times (-i\slashed{\Sigma}_{\ell})$, with 
\begin{align} \label{eq:S>ji}
	i\slashed{{S}}^>_{\ell_{ji}}&=(i\slashed{{S}}^R_{\ell_i})(-i\slashed{\Sigma}^T_{\ell})(i\slashed{{S}}^>_{\ell_j})+(i\slashed{{S}}^R_{\ell_i})(-i\slashed{\Sigma}^>_{\ell})(i\slashed{{S}}^R_{\ell_j})+(i\slashed{{S}}^<_{\ell_i})(-i\slashed{\Sigma}^>_{\ell})(i\slashed{{S}}^R_{\ell_j})
	\nonumber\\[0.2cm]
	&-(i\slashed{{S}}^R_{\ell_i})(-i\slashed{\Sigma}^>_{\ell})(i\slashed{{S}}^>_{\ell_j})-(i\slashed{{S}}^>_{\ell_i})(-i\slashed{\Sigma}^{\bar T}_{\ell})(i\slashed{{S}}^R_{\ell_j})\,.
\end{align}
Note that we have used the following relations:
\begin{align}
	(i\slashed{{S}}^R_{\ell_i})(-i\slashed{\Sigma}_{\ell})(i\slashed{{S}}^{<,>}_{\ell_j})&=(i\slashed{{S}}^{<,>}_{\ell_j})(-i\slashed{\Sigma}_{\ell})	(i\slashed{{S}}^R_{\ell_i})\,,\label{eq:R<>commute}
	\\[0.2cm]
	(i\slashed{{S}}^R_{\ell_i})(-i\slashed{\Sigma}_{\ell})(i\slashed{{S}}^R_{\ell_j})&=(i\slashed{{S}}^R_{\ell_j})(-i\slashed{\Sigma}_{\ell})(i\slashed{{S}}^R_{\ell_i})\,,\label{eq:RRcommute}
\end{align} 
since the resummed retarded propagator and Wightman functions depend on the same helicity operator $P_\pm$. 

In the weak washout regime where   the distribution functions of $\chi_1, \bar\chi_1$ can be neglected,  we can write down the  CP-violating source $\mathcal{S}_{\rm CP}$ from Eq.~\eqref{eq:0th-KB} as
\begin{align}\label{eq:S_CP-2}
	\mathcal{S}_{\rm CP}(p)=\frac{1}{2}\int \frac{d^4 p_{}}{(2\pi)^4}(2\pi)\delta(p^2-m_{\chi_1}^2) \text{Tr}[\slashed{K} \slashed{p}]\,,
\end{align}
where  $\slashed{K}$ is defined as $\slashed{K}(p_{})\equiv \theta(-p_{0})i\slashed{\Sigma}_{\chi_1}^>- \theta(p_{0})i\slashed{\Sigma}_{\chi_1}^<$ and is given by 
\begin{align}\label{eq:trace}
	\slashed{K}(p_{})&=-2\int \frac{d^4 p_\ell}{(2\pi)^4}\frac{d^4 p_\phi}{(2\pi)^4} (2\pi)^4\delta^4(p_{}-p_\ell+p_\phi) \sum_{i=1}^{5}\mathcal{I}_i\,,
\end{align}
with the five $\mathcal{I}_i$ functions from the difference of Eqs.~\eqref{eq:S>ji} and~\eqref{eq:tildeS<Lij}:
\begin{align}
	\mathcal{I}_1&=i\slashed{{S}}^R_{\ell_i} \left[Y_4^*\theta(-p_{ 0})e^{p_{\ell 0}/T}iG_\phi^<(-i\slashed{\Sigma}^T_{\ell})+Y_4\theta(p_{0})iG_\phi^>(-i\slashed{\Sigma}^T_{\ell})\right]i\slashed{{S}}^<_{\ell_j}\,,
	\\[0.2cm]
	\mathcal{I}_2&=-i\slashed{{S}}^R_{\ell_i}  \left[Y_4^*\theta(-p_{0})iG_\phi^< (-i\slashed{\Sigma}^>_{\ell})-Y_4\theta(p_{0})iG_\phi^>(-i\slashed{\Sigma}^<_{\ell})\right] i\slashed{{S}}^R_{\ell_j}\,,
	\\[0.2cm]
	\mathcal{I}_3&=-i\slashed{{S}}^<_{\ell_i}\left[Y_4^*\theta(-p_{0})iG_\phi^<(-i\slashed{\Sigma}^>_{\ell})-Y_4\theta(p_{0})iG_\phi^>(-i\slashed{\Sigma}^<_{\ell})\right]i\slashed{{S}}^R_{\ell_j}\,,
	\\[0.2cm]
	\mathcal{I}_4&=-i\slashed{{S}}^R_{\ell_i}\left[Y_4^*\theta(-p_{0})e^{p_{\ell 0}/T}iG_\phi^<(-i\slashed{\Sigma}^>_{\ell})-Y_4\theta(p_{0})e^{p_{\ell 0}/T} iG_\phi^>(-i\slashed{\Sigma}^<_{\ell})\right]i\slashed{{S}}^<_{\ell_j}\,,
	\\[0.2cm]
	\mathcal{I}_5&=-i\slashed{{S}}^<_{\ell_i}\left[Y_4^*\theta(-p_{0})e^{p_{\ell 0}/T}iG_\phi^<(-i\slashed{\Sigma}^{\bar T}_{\ell})+Y_4\theta(p_{0})iG_\phi^>(-i\slashed{\Sigma}^{\bar T}_{\ell})\right] i\slashed{{S}}^R_{\ell_j}\,.
\end{align}
Note that in Eq.~\eqref{eq:S_CP-2}, $m_{\chi_2}$ will not contribute in the Dirac trace since there is no chirality flip in the inner loop of Fig.~\ref{fig:lep-cuts}. Besides, $\text{Tr}[\slashed{K}m_{\chi_1}]=0$ since $\slashed{K}$ contains an odd number of Dirac $\gamma$-matrices. The factor of $e^{p_{\ell 0}/T}$ in $\mathcal{I}_1,\mathcal{I}_4, \mathcal{I}_5$ arises from the KMS relation  $\slashed{{S}}^>_{\ell}(p)=-e^{p_{0}/T}\slashed{S}_\ell^<(p)$, and we have defined  
\begin{align}
	Y_4\equiv y'_{i1}y^{\prime *}_{j1}y_{i2}^{\prime *}y'_{j2}\,.
\end{align}
$\mathcal{I}_3$ and $\mathcal{I}_5$ can be rewritten as a product of $i{S}^R_{\ell_i}\times i{S}^<_{\ell_j}$ similar to $\mathcal{I}_1$ and $\mathcal{I}_4$. To this end, we interchange the dummy indices $i,j$ and use $Y_4(i\leftrightharpoons j)=Y_4^*$ as well as Eq.~\eqref{eq:R<>commute}, resulting in 
\begin{align}
	\mathcal{I}_3&
	=-i\slashed{{S}}^R_{\ell_i}\left[Y_4\theta(-p_{0})iG_\phi^<(-i\slashed{\Sigma}^>_{\ell})-Y_4^*\theta(p_{0})iG_\phi^>(-i\slashed{\Sigma}^<_{\ell})\right] i\slashed{{S}}^<_{\ell_j}\,,\label{eq:I3re}
	\\[0.2cm]
	\mathcal{I}_5&= -i\slashed{{S}}^R_{\ell_i}\left[Y_4 \theta(-p_{0})e^{p_{\ell 0}/T}iG_\phi^<(-i\slashed{\Sigma}^{\bar T}_{\ell})+Y_4^*\theta(p_{0})iG_\phi^>(-i\slashed{\Sigma}^{\bar T}_{\ell})\right] i\slashed{{S}}^<_{\ell_j}\,.\label{eq:I5re}
\end{align}
After this performance, Eq.~\eqref{eq:trace} will only contain products of $i{S}^R_{\ell_i}\times i{S}^<_{\ell_j}$  and $i{S}^R_{\ell_i}\times i{S}^R_{\ell_j}$. 

The term $i{S}^R_{\ell_i}\times i{S}^<_{\ell_j}$ corresponds to making the lepton propagator  of flavor $j$ onshell, i.e., the decay  and inverse decay channels induced from the blue cutting line in Fig.~\ref{fig:lep-cuts}, while the term  $i{S}^R_{\ell_i}\times i{S}^R_{\ell_j}$ represents the scattering processes mediated by onshell  and offshell lepton propagators  induced  from the red cutting line. The onshell scattering contribution is crucial to   ensure that  no CP asymmetry can arise from   full thermal equilibrium, and hence to guarantee CPT theorem and unitarity.  This contribution also corresponds to the effect that should be carefully subtracted in the Boltzmann equation, known as the real-intermediate-state subtraction~\cite{Kolb:1979qa,Giudice:2003jh}.  In the following calculation, we will take into account the onshell scattering effect but   neglect the offshell contribution. It should be mentioned, however, in the strong washout regime, the offshell scattering contribution could be important to erase the CP asymmetry generated at earlier epochs.

To proceed, let us rewrite the CP-violating source in a compact form:
\begin{align}\label{eq:S_CP-3}
	\mathcal{S}_{\rm CP}(p)=-\int \frac{d^4 p_{}}{(2\pi)^4}\int \frac{d^4 p_{\ell}}{(2\pi)^4}\int \frac{d^4 p_{\phi}}{(2\pi)^4}(2\pi)^5\delta^4(p_{}-p_\ell+p_\phi)\delta(p^2-m_{\chi_1}^2) \sum_{i=1}^{5}\mathcal{S}_{\text{CP}i}\,,
\end{align}
where $\mathcal{S}_{\text{CP}i}\equiv \text{Tr}[\mathcal{I}_i \slashed{p}]$. For the moment, we take
the thermal distribution for leptons, but allow   nonthermal distributions for  scalar $\phi$ and fermion $\chi_2$. As will be shown below, the distributions of    $\phi$ and   $\chi_2$ in the inner loop of Fig.~\ref{fig:lep-cuts} determine if the forbidden CP asymmetry can arise. In particular, for  $\phi$ and   $\chi_2$  in full thermal equilibrium, there is no forbidden CP asymmetry of $\chi_1$. 
To parameterize the departure from thermal equilibrium, we define $f_i=f_i^{\rm eq}+\delta f_i$ for $i=\phi, \chi_2$, with the approximation $\delta f_i=\delta \bar f_{i}$. This treatment assumes that there are no large initial asymmetries of $\phi,\chi_2$. For example, both $\phi$ and $\chi_2$ are initially in thermal equilibrium ($T>m_\phi$) prior to the onset of the leptogenesis epoch ($T\lesssim m_\phi$).   Accordingly, we can write the propagators and self-energy amplitudes as
\begin{align}\label{eq:Gab}
	iG^{ab}_\phi (p)&=iG^{ab, \rm eq}_{\phi}(p)+i \delta G_\phi(p)\,,
	\\[0.2cm]
	i\slashed{S}^{ab}_{\chi_2} (p)&=i\slashed{S}^{ab, \rm eq}_{\chi_2}(p)+i \delta \slashed{S}^{}_{\chi_2}(p)\,,\label{eq:Sab}
	\\[0.2cm]
	i\slashed{\Sigma}_\ell^{ab}(p)&=i\slashed{\Sigma}_{\ell}^{ab,\rm eq}(p)+i\delta \slashed{\Sigma}_\ell^{ab}(p)\,,\label{eq:Sigmaab}
\end{align}
where $a,b=\pm$, and 
\begin{align}
	i \delta G_\phi(p)&=2\pi \delta(p^2-m_\phi^2)\delta f_\phi(|p_0|)\,,
	\\[0.2cm]
	i \delta \slashed S_{\chi_2}(p)&=-2\pi \delta(p^2-m_{\chi_2}^2)(\slashed{p}+m_{\chi_2})\delta f_{\chi_2}(|p_0|)\,,
	\\[0.2cm]
	i \slashed{\Sigma}_{\ell}^{ab,\rm eq}(p_\ell)&= \int \frac{d^4q}{(2\pi)^4} \frac{d^4q_\phi}{(2\pi)^4}(2\pi)^4 \delta^4(q-q_\phi-p_\ell) P_R i\slashed{S}_{\chi_2}^{ab,\rm eq}P_L iG_\phi^{ba,\rm eq}\,,
	\\[0.2cm]
	i\delta \slashed{\Sigma}_{\ell}^{ab}(p_\ell)&= \int \frac{d^4q}{(2\pi)^4} \frac{d^4q_\phi}{(2\pi)^4}(2\pi)^4 \delta^4(q-q_\phi-p_\ell)  \left(P_R i\slashed{S}_{\chi_2}^{ab,\rm eq}P_L i\delta G_\phi+P_Ri\delta \slashed{S}_{\chi_2}^{}P_L i G^{ba,\rm eq}_\phi \right)\,,
\end{align}
with $q,q_\phi$ the inner loop momenta of $\chi_2$ and $\phi$.  Note that we have only kept the linear deviation of thermal distributions in  $i\delta \slashed{\Sigma}_{\ell}^{ab}$.

There are two terms in each $\mathcal{S}_{\text{CP}i}$, corresponding to positive and negative frequencies of $\chi_1$. Note that the presence of $iG_\phi^{<,>}$ from the outer loop dictates the kinetic threshold via 
$\delta(p_\phi^2-m_\phi^2)$. For $m_{\chi_1}^2+2\tilde{m}_i^2<m_\phi^2$ we obtain     two kinetic regions:
\begin{align}\label{eq:kinetic1}
	\theta(-p_0)\theta(p_{\ell 0})\,,\quad \theta(p_0)\theta(-p_{\ell 0})\,.
\end{align}
Using these regions and the symmetries of the full momentum integration, we can simplify the $\theta(-p_0)$ and $\theta(p_0)$ terms in each $\mathcal{S}_{\text{CP}i}$. Starting with  $\mathcal{S}_{\text{CP}1}$, we see that  for the $\theta(-p_0)$ term, only the positive lepton frequency from $i\slashed{S}_{\ell_j}^<$ is selected, and the resonance would  appear in  the $P_- $ component of $\slashed S^R_{\ell_i}$. Analogously, the negative lepton frequency from $i\slashed{S}_{\ell_j}^<$ is selected in the $\theta(p_0)$ term and the resonance would  appear in  the $P_+ $ component of $\slashed S^R_{\ell_i}$. Consequently, we obtain the $\theta(-p_0)$  and     $\theta(p_0)$ terms of $\mathcal{S}_{\text{CP}1}$:
\begin{align}
	\mathcal{S}_{\text{CP}1}[\theta(-p_0)]&= \frac{-i\pi  Y_4^*}{2(\tilde{m}_j^2-\tilde{m}_i^2)}\theta(-p_0)e^{p_{\ell 0}/T}iG_{\phi}^{<}(-i \Sigma^T_{\ell})\text{sign}(p_{\ell 0})f_\ell(p_{\ell 0})\frac{\delta (p_{\ell 0}-\omega_j)}{\omega_j}\text{Tr}\left[P_L\slashed{p}_\ell P_R\slashed{q}P_L \slashed{p}_{\ell}P_R\slashed{p}\right],\label{eq:SCP1-1}
	\\[0.2cm]
	\mathcal{S}_{\text{CP}1}[\theta(p_0)]&= \frac{-i\pi  Y_4}{2(\tilde{m}_j^2-\tilde{m}_i^2)}\theta(p_0)iG_{\phi}^{>}(-i \Sigma^T_{\ell})\text{sign}(p_{\ell 0})f_\ell(p_{\ell 0})\frac{\delta (p_{\ell 0}+\omega_j)}{\omega_j}\text{Tr}\left[P_L\slashed{p}_\ell P_R\slashed{q}P_L \slashed{p}_{\ell}P_R\slashed{p}\right],\label{eq:SCP1-2}
\end{align}
where we have factored out the Dirac matrices from $-i \slashed\Sigma^T_{\ell}$ via $-i \slashed\Sigma^T_{\ell}=-i \Sigma^T_{\ell}\times (P_R\slashed{q}P_L)$.
By applying  $p_i\to -p_i$   for all momenta to Eq.~\eqref{eq:SCP1-1} and using  
\begin{align}\label{eq:Z2-symmetry}
	G_\phi^<(-p)=G_\phi^>(p)\,,\quad \Sigma^T_{\ell}(-p)= \Sigma^T_{\ell}(p)\,, \quad e^{-p_{\ell 0}/T}f_\ell(-p_{\ell 0})=f_\ell(p_{\ell 0})\,,
\end{align}  
we obtain a simplified $\mathcal{S}_{\text{CP}1}$:
\begin{align}
	\mathcal{S}_{\text{CP}1}&=\frac{\pi \text{Im}(Y_4)}{\tilde{m}_j^2-\tilde{m}_i^2 }\theta(p_0) iG_{\phi}^{>}(-i \Sigma^T_{\ell})\text{sign}(p_{\ell 0})f_\ell(p_{\ell 0})\frac{\delta (p_{\ell 0}+\omega_j)}{\omega_j}\text{Tr}\left[P_L\slashed{p}_\ell P_R\slashed{q}P_L \slashed{p}_{\ell}P_R\slashed{p}\right].
\end{align}
The above analysis can be directly applied to $	\mathcal{S}_{\text{CP}5}$ with $\mathcal{I}_5$ given by Eq.~\eqref{eq:I5re}. We obtain the same  expression for  $\mathcal{S}_{\text{CP}5}$  except for the $\Sigma_\ell$ dependence:
\begin{align}
	\mathcal{S}_{\text{CP}5}&=\frac{\pi \text{Im}(Y_4)}{\tilde{m}_j^2-\tilde{m}_i^2 }\theta(p_0) iG_{\phi}^{>}(-i \Sigma^{\bar T}_{\ell})\text{sign}(p_{\ell 0})f_\ell(p_{\ell 0})\frac{\delta (p_{\ell 0}+\omega_j)}{\omega_j}\text{Tr}\left[P_L\slashed{p}_\ell P_R\slashed{q}P_L \slashed{p}_{\ell}P_R\slashed{p}\right].
\end{align}
Next, let us turn to $\mathcal{S}_{\text{CP}3}$ with $\mathcal{I}_3$ given by Eq.~\eqref{eq:I3re}. We can analogously write the $\theta(-p_0)$ and $\theta(p_0)$ terms as 
\begin{align}
	\mathcal{S}_{\text{CP}3}[\theta(-p_0)]&=\frac{i \pi Y_4}{2(\tilde{m}_j^2-\tilde{m}_i^2)}\theta(-p_0)iG_\phi^< (-i\Sigma_\ell^>)\text{sign}(p_{\ell 0})f_{\ell}(p_{\ell 0})\frac{\delta(p_{\ell 0}-\omega_j)}{\omega_j}\text{Tr}\left[P_L\slashed{p}_\ell P_R\slashed{q}P_L \slashed{p}_{\ell}P_R\slashed{p}\right],\label{eq:SCP3-1}
	\\[0.2cm]
	\mathcal{S}_{\text{CP}3}[\theta(p_0)]&=\frac{-i\pi  Y_4^*}{2(\tilde{m}_j^2-\tilde{m}_i^2)}\theta(p_0)iG_\phi^> (-i\Sigma_\ell^<)\text{sign}(p_{\ell 0})f_{\ell}(p_{\ell 0})\frac{\delta(p_{\ell 0}+\omega_j)}{\omega_j}\text{Tr}\left[P_L\slashed{p}_\ell P_R\slashed{q}P_L \slashed{p}_{\ell}P_R\slashed{p}\right].\label{eq:SCP3-2}
\end{align}
By applying  $p_i\to -p_i$ to Eq.~\eqref{eq:SCP3-1} and using Eq.~\eqref{eq:Z2-symmetry} as well as
\begin{align}
	i\Sigma_\ell^>(-p)=i\Sigma_\ell^<(p)\,,\quad   f_\ell(-p_{\ell 0})+f_\ell(p_{\ell 0})=1\,, 
\end{align}
we arrive at  
\begin{align}
	\mathcal{S}_{\text{CP}3}&=\frac{-i\pi  Y_4}{2(\tilde{m}_j^2-\tilde{m}_i^2)}\theta(p_0)iG_\phi^> (-i\Sigma_\ell^<)\text{sign}(p_{\ell 0})\frac{\delta(p_{\ell 0}+\omega_j)}{\omega_j}\text{Tr}\left[P_L\slashed{p}_\ell P_R\slashed{q}P_L \slashed{p}_{\ell}P_R\slashed{p}\right]
	\nonumber\\[0.2cm]
	&-\frac{\pi \text{Im}(Y_4)}{\tilde{m}_j^2-\tilde{m}_i^2}\theta(p_0)iG_\phi^> (-i\Sigma_\ell^<)\text{sign}(p_{\ell 0})f_{\ell}(p_{\ell 0})\frac{\delta(p_{\ell 0}+\omega_j)}{\omega_j}\text{Tr}\left[P_L\slashed{p}_\ell P_R\slashed{q}P_L \slashed{p}_{\ell}P_R\slashed{p}\right].
\end{align}
Using the same trick, we  find that 
\begin{align}
	\mathcal{S}_{\text{CP}4}=\mathcal{S}_{\text{CP}3}\,. 
\end{align}

The simplified $\mathcal{S}_{\text{CP}1},\mathcal{S}_{\text{CP}3},\mathcal{S}_{\text{CP}4}, \mathcal{S}_{\text{CP}5}$ all have the same correspondence to the decay and inverse decay processes, which are induced by the blue cutting line shown in Fig.~\ref{fig:lep-cuts}. The remaining $\mathcal{S}_{\text{CP}2}$ has a different structure, which is  characterized by  the product of  two retarded propagators. This term has two scattering channels mediated by resummed leptons of flavor $i$ and $j$, which corresponds to onshell $\slashed{S}^R_{\ell_i}$ and onshell $\slashed{S}^R_{\ell_j}$ respectively.  Using Eq.~\eqref{eq:tildeS^R-onshell} for flavor $i$ and $j$ separately  and then interchanging the dummy indices $i,j$ for onshell flavor $i$, we arrive at
\begin{align}
	\mathcal{S}_{\text{CP}2}=\frac{ i\pi (Y_4^*+Y_4)}{2(\tilde{m}_j^2-\tilde{m}_i^2)}\theta(p_0)iG_\phi^> (-i\Sigma_\ell^<)\text{sign}(p_{\ell 0})\frac{\delta(p_{\ell 0}+\omega_j)}{\omega_j}\text{Tr}\left[P_L\slashed{p}_\ell P_R\slashed{q}P_L \slashed{p}_{\ell}P_R\slashed{p}\right].
\end{align}
We see that the  $Y_4$ dependence of $\mathcal{S}_{\text{CP}2}$ is to be combined with that of $\mathcal{S}_{\text{CP}3}, \mathcal{S}_{\text{CP}4}$, leading to a CP-violating source that depends only on the imaginary part of Yukawa couplings: $\text{Im}(Y_4)$. 

Assembling the five $\mathcal{S}_{\text{CP}i}$, we arrive at
\begin{align}\label{eq:S_CP-4}
	\mathcal{S}_{\rm CP}(p)&=\frac{\pi \text{Im}(Y_4)}{\tilde{m}_j^2-\tilde{m}_i^2}\int \frac{d^4 p_{}}{(2\pi)^4}\int \frac{d^4 p_{\ell}}{(2\pi)^4}\int \frac{d^4 p_{\phi}}{(2\pi)^4}(2\pi)^5\delta^4(p_{}-p_\ell+p_\phi)\delta(p^2-m_{\chi_1}^2)\theta(p_0)\text{sign}(p_{\ell 0})
	\nonumber\\[0.2cm]
	&\times iG_\phi^>\left\{\left(i\Sigma_{\ell}^>-i\Sigma_{\ell}^<\right) f_{\ell}(p_{\ell 0})+ i\Sigma_{\ell}^<\right\}\frac{\delta(p_{\ell 0}+\omega_j)}{\omega_j}\text{Tr}\left[P_L\slashed{p}_\ell P_R\slashed{q}P_L \slashed{p}_{\ell}P_R\slashed{p}\right].
\end{align}
We see that for thermal $\phi$ and $\chi_2$, the KMS relation holds $\Sigma_\ell^>(p)=-e^{p_0/T}\Sigma_{\ell}^<(p)$. Consequently, the terms in the curly bracket sum to zero:
\begin{align}
	\left[i\Sigma_{\ell}^{>,\rm eq}(p_{\ell})- i\Sigma_{\ell}^{<,\rm eq}(p_{\ell})\right] f_{\ell}(p_{\ell 0})+ i\Sigma_{\ell}^{<,\rm eq}(p_{\ell})=0\,.\label{eq:KMS}
\end{align} 
Therefore, a nonthermal $\chi_1$ is not sufficient to generate a nonzero forbidden CP asymmetry and at least $\phi$ or $\chi_2$ should additionally be  out of equilibrium during the leptogenesis epoch.    

For $m_\phi\gg m_{\chi_2}$, the scalar starts to significantly depart from thermal equilibrium at $T<m_\phi$. In this epoch,  $\chi_2$ is still relativistic, such that scattering between $\chi_2$ and the SM plasma is strong enough to keep $\chi_2$ in thermal equilibrium, provided the Yukawa coupling is not too small. Therefore, we will take the thermal distribution for $\chi_2$ to evaluate Eq.~\eqref{eq:S_CP-4}. 
Substituting Eqs.~\eqref{eq:Gab} and~\eqref{eq:Sigmaab} into Eq.~\eqref{eq:S_CP-4} and using Eq.~\eqref{eq:KMS},  we obtain the $iG_\phi^>\times \{...\}$ term of Eq.~\eqref{eq:S_CP-4} at linear order of nonthermal distribution functions:
\begin{align}
	iG_\phi^{>,\rm eq}\left\{(i\delta \Sigma_\ell^>-i\delta \Sigma_\ell^<)f_{\ell}(p_{\ell 0})+i\delta \Sigma_\ell^{<}\right\},
\end{align}
where the curly bracket term reads
\begin{align}\label{eq:bracket}
	\{...\}=\int \frac{d^4 q_\phi}{(2\pi)^2}d^4 q  
	\delta^4(q-q_\phi-p_\ell)\delta(q^2-m_{\chi_2}^2)\delta(q_\phi^2-m_\phi^2)\delta f_\phi(|q_{\phi 0}|)\left[\text{sign}(q_0)f_\ell(p_{\ell 0})-f_{\chi_2}(|q_0|)+\theta(-q_0)\right].
\end{align}
The Dirac $\delta$-functions dictate the kinetic regions. From $\delta(q^2-m_{\chi_2}^2)$ and  $\delta(q_\phi^2-m_{\phi}^2)$, together with the second kinetic region given in Eq.~\eqref{eq:kinetic1}, we obtain 
\begin{align}\label{eq:kinetic2}
	q_{\phi 0}>0, \quad q_0>0\,,
\end{align}
and  Eq.~\eqref{eq:bracket} reduces to
\begin{align}
	\{...\}=\frac{1}{8\pi p_\ell}\int^\infty_{\frac{m_\phi^2}{4p_\ell}+p_\ell} dE'_\phi \delta_\phi(E'_\phi)\left[1-f_{\chi_2}(E'_\phi-p_\ell)-f_\ell(p_\ell)\right],
\end{align}
where  $E'_\phi=|q_{\phi 0}|$,  the lower integration limit comes from the angular integration with $\delta(m_\phi^2-2E'_\phi p_\ell+2p_\ell q_\phi \cos\theta)$, and  we have  used $f_{\ell}(p_{\ell 0})=1-f_{\ell}(-p_{\ell 0})$ for $p_{\ell 0}=-\omega_j$ with the  approximation $ \omega_j\approx |\vec p_\ell|\equiv p_\ell$. 

Integrating over $d^4 p$ via $\delta^4(p-p_\ell+p_\phi)$ in Eq.~\eqref{eq:S_CP-4}, $dp_{\ell 0}$ via $\delta(p_{\ell 0}+\omega_j)$, and $dp_{\phi 0}$ via $\delta(p_{\phi 0}+E_\phi)$ (from $iG_\phi^{>,\rm eq}$), we finally arrive at the result presented in the Letter:
\begin{align}
	S_{\rm CP}=-\frac{\text{Im}(Y_4)m_\phi^4}{256\pi^4 (\tilde{m}_j^2-\tilde{m}_i^2)}\int^\infty_0 \frac{dp_\ell}{p_\ell}\int^\infty_{\frac{m_\phi^2}{4p_\ell}+p_\ell} dE_\phi \int^\infty_{\frac{m_\phi^2}{4p_\ell}+p_\ell} dE'_\phi f_\phi(E_\phi)\delta f_\phi(E'_\phi)\left[1-f_{\chi_2}(E'_\phi-p_\ell)-f_\ell(p_\ell)\right],
\end{align}
where $E_\phi=|p_{\phi 0}|$, $\text{Tr}[P_L\slashed{p}_\ell P_R\slashed{q}P_L \slashed{p}_{\ell}P_R\slashed{p}]=4(p\cdot p_\ell)(q\cdot p_\ell)-2p_\ell^2(p\cdot q)\approx m_\phi^4$ was used in the limit $m_\phi^2\gg \tilde{m}_j^2, m_{\chi_1}^2, m_{\chi_2}^2$,   and   the lower integration limit  of $E_\phi$ comes from the angular integration with $\delta(m_\phi^2-2E_\phi p_\ell+2p_\ell p_\phi \cos\theta)$. It is worth mentioning that the scaling $1/p_\ell$ in the $p_\ell$ integration will not lead to  IR divergence.  This  can be seen by taking $p_\ell\to 0$ in the lower $E_\phi, E'_\phi$ integration limits, which  makes the integration turn to zero.

\end{document}